\documentclass[lettersize,journal]{IEEEtran}

\usepackage{amsfonts}
\usepackage{amsmath} 
\usepackage{array}
\usepackage{cite}
\usepackage{algorithm}  
\usepackage{algorithmic}
\usepackage{verbatim} 
\usepackage{booktabs}
\usepackage[caption=false,font=normalsize,labelfont=sf,textfont=sf]{subfig}
\usepackage{textcomp}
\usepackage{stfloats}
\usepackage{float}
\usepackage{url}
\usepackage{verbatim}
\usepackage{graphicx}
\usepackage{amssymb}
\usepackage{color}
\usepackage{graphics}
\usepackage{epstopdf}
\usepackage{bbm}
\usepackage[T1]{fontenc}% optional T1 font encoding
\usepackage{balance}
\usepackage{hyperref}
\begin{document}

\title{GNN Based Joint Beamforming Design for Extremely Large-Scale RIS Assisted Near-Field ISAC Systems}

\author{Jiahao Chen,~\IEEEmembership{Student Member, IEEE}, Feng Wang,~\IEEEmembership{Member, IEEE}, Guojun Han,~\IEEEmembership{Senior Member, IEEE}, \\ Xin Wang,~\IEEEmembership{Fellow, IEEE}, and Vincent K. N. Lau,~\IEEEmembership{Fellow, IEEE}

\thanks{This paper was presented in part at the IEEE/CIC International Conference on Communications in China (ICCC), Shanghai China, August 10--13\cite{Conf_version}.}

\thanks{J. Chen, F. Wang, and G. Han are with the School of Information Engineering, Guangdong University of Technology, Guangzhou 510006, China (e-mail: fengwang13@gdut.edu.cn).}

\thanks{X. Wang is with the Key Laboratory for Information Science of Electromagnetic Waves (MoE), College of Future Information Technology, Fudan University, Shanghai 200438, China (e-mail: xwang11@fudan.edu.cn).}

\thanks{V. K. N. Lau is with the Department of Electronic and Computer Engineering, The Hong Kong University of Science and Technology, Hong Kong (e-mail: eeknlau@ust.hk).}

%\vspace{-0.5cm}
}

\maketitle
\begin{abstract}
This paper investigates an extremely large-scale reconfigurable intelligent surface (XL-RIS) assisted near-field integrated sensing and communication (ISAC) system, where a multi-antenna base station (BS) simultaneously sends unicast data to multiple single-antenna communication users (CUs) and senses multiple targets (TGTs). The BS, CUs and TGTs are \emph{all} assumed to be located in the near-field region of the XL-RIS. We aim to maximize the weighted sum rate (WSR) of all CUs, subject to the sensing beampattern gain constraint for each TGT, the transmit power constraint for the BS, and the unit modulus constraints on the XL-RIS phase shift. First, we develop a fractional programming (FP) based block coordinate descent (BCD) algorithm to obtain a locally optimal solution for such a non-convex joint design problem. Secondly, to address the high-dimensional spatial correlations and scalability of the XL-RIS near-field channels, we propose a customized graph neural network (GNN) scheme to generate the BS transmit beamforming variables and the XL-RIS reflecting coefficient vector for ISAC, where the near-field ISAC system is modeled as a heterogeneous graph comprising XL-RIS/CU/TGT nodes. The proposed GNN scheme can effectively learn the near-field channel state information (CSI) features, in which the message passing mechanism is employed to exchange CSI among these directly connected nodes in the graph. Furthermore, each XL-RIS/CU/TGT node maintains a feature vector for mapping to the BS transmit beamforming variables or the XL-RIS reflecting coefficient vector. Numerical results show that the proposed GNN-based beamforming design scheme achieves a better performance than the existing baselines, in terms of computational efficiency, feasibility, robustness, and the ability of generalization.
\end{abstract}


\section{Introduction}

\IEEEPARstart{S}{ixth-generation} (6G) wireless networks have been envisioned as critical enablers for many emerging applications calling for integrated sensing and communication (ISAC). The resources of communication and sensing can be shared in ISAC systems, which allow communication and radar sensing functionalities to coexist on a hardware platform and share the same transmit waveform, thereby inherently boosting the efficiency of energy, spectrum, and hardware. However, the growing scarcity of ISAC resources is likely to constrain the capacity of 6G wireless networks\cite{b00,b01,b011}.

Accordingly, the reconfigurable intelligent surface (RIS) has been regarded as a promising solution to meet the demands of 6G wireless networks\cite{b02,b03}. Assisted by RIS, ISAC is expected to achieve significant performance improvements in sensing and communication\cite{b0}. In addition, with the integration of extremely large-scale antenna arrays and the utilization of higher frequencies such as millimeter wave (mmWave) and terahertz (THz), it becomes more important to address the near-field effects as the near-field region expands\cite{b1}. Consequently, realizing such near-field ISAC systems poses critical challenges, including maintaining real-time sensing and low-latency communication services.

\subsection{Related Work}
The integration of RIS into ISAC has drawn great attention for its ability to enhance both communication rates and sensing accuracy by reconfiguring the wireless propagation environment. A crucial challenge in ISAC is the joint beamforming design, which typically involves non-convex optimization problems with coupled variables. Conventionally, the optimization theory based algorithms including several techniques, e.g., fractional programming (FP), successive convex approximation (SCA), and block coordinate descent (BCD), have been widely employed to address the challenge. For instance, the works in \cite{b101,b16,b102} have utilized the aforementioned methods to maximize the minimum sensing beampattern among multiple targets (TGTs) and the weighted sum rate (WSR) of communication users (CUs), respectively.

However, these optimization theory based approaches face severe limitations in extremely large-scale reconfigurable intelligent surface (XL-RIS) assisted near-field ISAC systems, which typically entail more runtime and fail to meet the real-time communication and sensing requirements as the number of XL-RIS elements increases. Furthermore, the complex coupled channel models in near-field region may exacerbate the convergence difficulties of traditional solvers.

To overcome the high computational overhead of traditional optimization, deep learning (DL)-based algorithms have emerged as the alternative to address various problems, such as channel estimation\cite{b2} and resource allocation\cite{b3}. Under imperfect channel state information (CSI), the authors in \cite{b42} investigated a heterogeneous multi-agent deep deterministic policy gradient (DDPG) based multi-receiver communication system enabled by movable antennas. The work in \cite{b43} studied the problem of learning based beamforming schemes that leverage user pilots to maximize the WSR in multi-user RIS systems while accommodating different service priorities and accounting for fairness via user weights based on a novel hypernetwork beamforming (HNB) framework. The authors in \cite{b44} studied a decentralized deep neural network (DNN) task partitioning and offloading control optimization problem for a multi-access edge computing system powered by renewable energy sources. The work \cite{b5} employed the DDPG neural networks to develop a joint design of the transmit beamforming matrix of BS and the phase shift matrix of RIS. The work \cite{b6} addressed a sequential task offloading problem for a multi-user MEC system by using DNN. Despite the promising performance of these DL models in several scenarios, they still suffer from a lack of generalization and scalability. Specifically, these DL models are structurally constrained by fixed-dimension inputs, meaning that a model trained for a specific number of CUs or TGTs must be retrained from scratch if the ISAC system topology changes. This limitation renders them inefficient for dynamic ISAC scenarios where the number of CUs or TGTs varies rapidly. 

 To address the generalization and scalability issues of conventional DL due to the fixed-dimension inputs, the graph neural network (GNN) has been introduced to ISAC systems. GNN can model the wireless network as a graph, exploiting the permutation invariance of the network topology, which allows GNN to generalize well to ISAC systems with varying numbers of CUs and TGTs without retraining.

This paper proposes to use the GNN based approach of the optimization for the XL-RIS assisted near-field ISAC systems. GNN has the advantage in exploiting the graph topology of wireless networks. The proposed approach is inspired by \cite{b4,b41,b60,b600}. In particular, \cite{b4} presents an RIS-assisted multi-user multiple-input single-output (MISO) communication system by utilizing GNN under the BS's transmit power constraint. Furthermore, the work in \cite{b41} proposes the pinching antenna systems based on GNN to learn transmit beamforming, and shows that the GNN-based method can achieve a high spectral efficiency with low inference complexity. The authors in \cite{b60} model the multiple-user MISO network as a graph, while directly mapping the CSI to beamforming vectors. Recently, the work in \cite{b600} develops a coordinated beamforming scheme for a multi-cell ISAC system in far-field region.

\subsection{Main Contribution}
Unlike existing far-field ISAC works, this paper proposes a GNN learning based beamforming design for near-field ISAC systems assisted by an XL-RIS, where the BS and multiple CUs/TGTs are located in the near-field region of the XL-RIS. We aim to maximize the achievable WSR by jointly optimizing the BS transmit beamforming vectors, sensing covariance matrix, and reflecting coefficients of XL-RIS, subject to the BS transmit power constraint, the sensing beampattern gain constraints, and the modulus constraints on the XL-RIS phase shift. Compared to the classic neural architectures (e.g., DNN/CNN), the proposed GNN-based solution exhibits a better generalizability in dynamical ISAC systems\cite{b61}.

Our main contributions are summarized as follows.

\begin{itemize}
    \item First, we develop a FP based BCD algorithm to obtain a local-optimal solution of the non-convex constrained WSR maximization problem. Based on the closed-form FP reformulation, the WSR problem is equivalently transformed as a biconvex problem. Leveraging BCD method, such biconvex problem is further decomposed into two convex subproblems, in which the BS's ISAC beamforming design variables and the XL-RIS reflecting coefficients are efficiently optimized in an alternating order.

    \item Second, we propose a novel GNN-based beamforming design framework for the XL-RIS assisted near-field ISAC systems, where the ISAC system is modeled as a heterogeneous graph and the XL-RIS/CUs/TGTs are defined as distinct nodes. All nodes in the graph can effectively capture the complex coupling between communication and sensing in the near-field region based on the high-fidelity near-field channel models established in this paper. The CU/TGT nodes compute the representation feature vectors by using their own near-field CSI, while the XL-RIS node constructs the representation feature vectors by processing global CSI to fulfill the requirement of simultaneously serving both communication and sensing.
    
    \item Third, the representation feature vectors of all nodes in the graph generated by near-field CSI are mapped to BS transmit beamforming variables for downlink communication/sensing and the XL-RIS reflecting coefficients. Unlike the DNN model, this graph-based architecture exploits the permutation invariance of the graph model, thereby allowing a superb generalizability to systems with varying numbers of CUs and TGTs in the near-field region of the XL-RIS.
    
    \item Finally, extensive simulation results demonstrate the effectiveness of the proposed design algorithms in terms of computational efficiency, feasibility, robustness, and the ability of generalization. Especially, the proposed GNN based beamforming design framework achieves better performance with significantly lower computational latency compared to the BCD-based algorithm, and shows the robustness against CSI imperfections in the near-field scenarios.
\end{itemize}

\subsection{Organization of the Paper and Notations}

The rest of the paper is organized as follows. Section \ref{2} introduces the XL-RIS assisted near-field ISAC system model and problem formulation. Section \ref{3} provides the proposed BCD method. Section \ref{4} presents the proposed GNN-based solution. Section \ref{5} provides numerical results to show the performance of the proposed scheme. Finally, we conclude the paper in Section \ref{6}.

The notations used in this paper are listed as follows. $\mathbb{E}\{\cdot\}$ denotes statistical expectation. ${\cal CN}(\mu,\sigma^2)$ denotes the circularly symmetric complex Gaussian (CSCG) distribution with mean $\mu$ and variance $\sigma^2$. $\boldsymbol{I}_N$ and $\mathbf{1}_N$ denote the identity matrix of size $N\times N$ and identity vector of size $N\times 1$, respectively. For any general matrix $\boldsymbol{A}$, $[\boldsymbol{A}]_{i,j}$ denotes the $i$-th row and $j$-th column element, and $A^*$, $A^T$ and $A^H$ denote the conjugate, the transpose and the conjugate transpose of $\boldsymbol{A}$, respectively. For any vector $\boldsymbol{a}$ (all vectors in this paper are column vectors), $\|\boldsymbol{a}\|$ denotes the Euclidean norm, $[\boldsymbol{a}]_{i}$ denotes the $i$-th element. $|x|$ denotes the absolute value of a complex number $x$, and $(x)^+=\max(x,0)$. $\Re\{\cdot\}$ and $\Im\{\cdot\}$ denote the real part and imaginary parts of a complex quantity (including matrix, vector, and complex number), respectively. Finally, the notation $\otimes$ denotes the Kronecker product operator.

\begin{figure}
    \centering
    \includegraphics[width=3.4in]{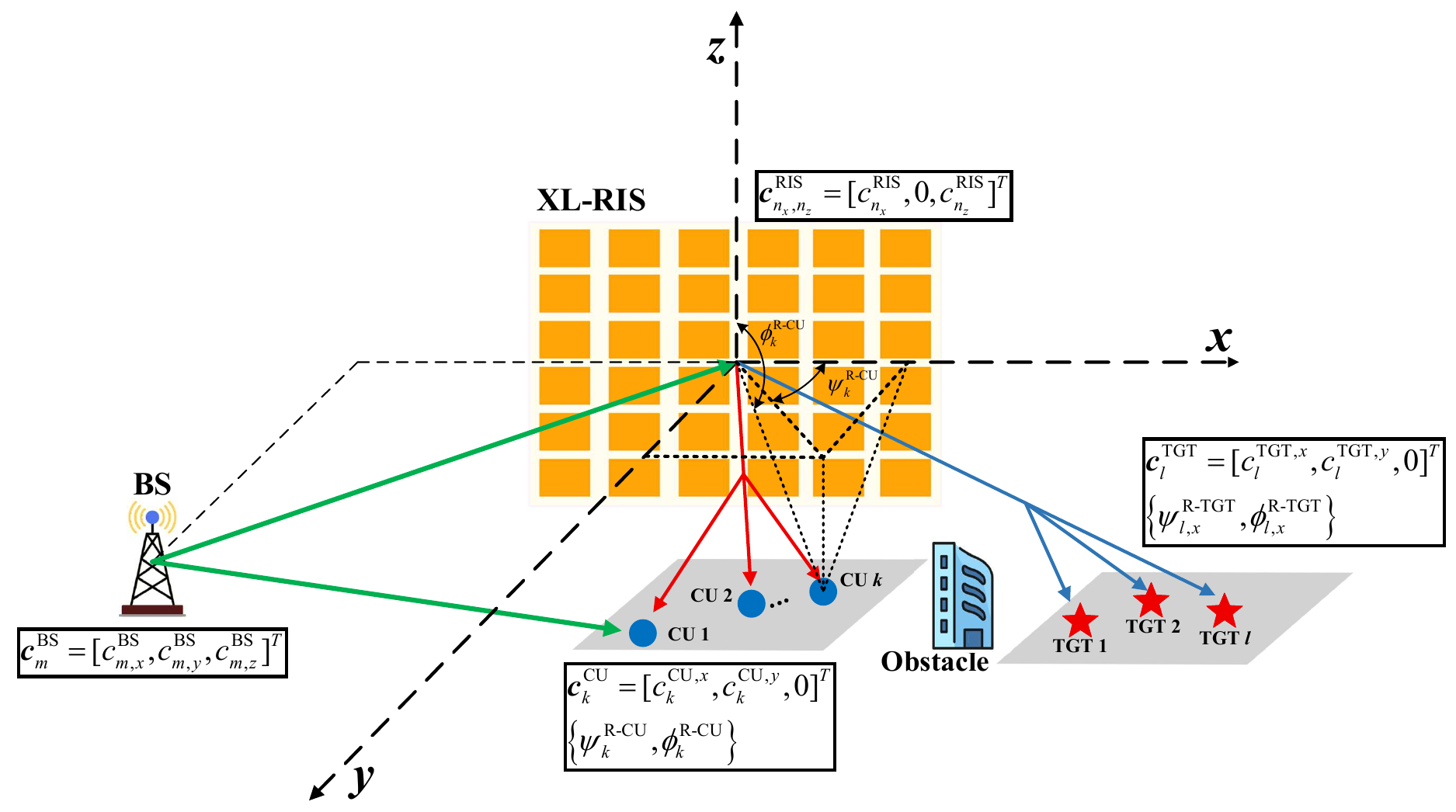}
    \caption{The XL-RIS assisted near-field ISAC system model, where the BS/CUs/TGTs are located in the near-field of XL-RIS.}\label{fig1}
\end{figure}
% \begin{figure}
%     \centering
%     \includegraphics[width=3in]{fig/simu1.pdf}
%     \caption{Simulation setup of the RIS assisted near-field ISAC systems.}\label{simu1}
% \end{figure}

\section{System Model and Problem Formulation}\label{2}

\subsection{System Model}
As shown in Fig.~\ref{fig1}, we consider an XL-RIS assisted near-field ISAC system, where a multi-antenna BS is responsible for simultaneously sending unicast data to a set ${\cal K}\triangleq\{1,2,...,K\}$ of $K$ single-antenna CUs in the downlink channels and sensing a set ${\cal L}\triangleq\{1,2,...,L\}$ of $L$ pointed-TGTs of interest. Note that due to the large scale of the XL-RIS, the BS and CUs/TGTs are assumed to be located in the near field region of the XL-RIS. In particular, the BS is equipped with a uniform linear array (ULA) of $M$ antennas, and the XL-RIS is equipped with a uniform planar array (UPA) of $N=N_xN_z$ reflecting elements, where $N_x=2\tilde{N}_x+1$ and $N_z=2\tilde{N}_z+1$ denote the number of reflecting elements of the XL-RIS along $x$-axis and $z$-axis, respectively. For both BS and XL-RIS, the antenna spacing is assumed to be $d=\frac{\lambda}{2}$, where $\lambda$ is the carrier wavelength of the BS's transmit signal for ISAC.

Note that in the absence of XL-RIS, the boundary between near-field and far-field is typically determined by the Rayleigh distance\cite{b8}, i.e., $r_R=\frac{2D^2}{\lambda}$, where $D=Md$ is the diameter of the BS's transmit antenna aperture. By contrast, in the case of XL-RIS presence, it is required to consider the cascaded channels from the BS to CU (or TGT) via the XL-RIS's reflection. The near-field range of the XL-RIS under consideration is given by 
$ r_h<\frac{2 D_{\text{RIS}}^2}{\lambda}$, where $D_{\text{RIS}}=\sqrt{(N_xd)^2+(N_zd)^2}$ denotes the effective diameter of the planar XL-RIS aperture, $r_h$ denotes the harmonic mean of the BS $\to$ XL-RIS distance and the XL-RIS $\to$ CU/TGT distance\cite{b9}. Furthermore, we consider that the systems operate in the Fresnel region of the near field ($1.2D_{\text{RIS}}<r_h<\frac{2 D_{\text{RIS}}^2}{\lambda}$). %Accordingly, the channel gain of each link between the BS antenna (or XL-RIS element) and the CU (or TGT) is assumed to be approximately identical \cite{b10}.

Denote by $s_k\in\mathbb{C}$ and $\boldsymbol{w}_k\in\mathbb{C}^{M\times 1}$ the information symbol and the transmit beamforming vector for each CU $k\in{\cal K}$, respectively. Without loss of generality, the information symbols $\{s_k\}_{k\in{\cal K}}$ are assumed to be zero-mean independent and identical distributed (i.i.d.) random variables with unit variance, i.e., $\mathbb{E}\{|s_k|^2\}=1$, $\forall k\in{\cal K}$. The equivalent baseband ISAC signal $\boldsymbol{x}\in\mathbb{C}^{M\times1}$ of the BS is modeled as\cite{b101}
 \begin{equation}\label{eq.x}
    \boldsymbol{x} = \sum_{k=1}^K\boldsymbol{w}_k {s}_k + \boldsymbol{x}_0,
 \end{equation}
 where $\boldsymbol{x}_0\in\mathbb{C}^{M\times1}$ denotes the BS's dedicated sensing symbol. Based on \eqref{eq.x}, the average transmit power of the BS is expressed as
 \begin{equation}
      \mathbb{E}[\|\boldsymbol{x}\|^2] 
      = \text{tr}(\boldsymbol{W}^H\boldsymbol{W})+\text{tr}(\boldsymbol{R}_0),
 \end{equation}
where $\boldsymbol{W}\triangleq[ \boldsymbol{w}_1,...,\boldsymbol{w}_K]\in{ \mathbb{C}}^{M\times K}$ denotes the BS communication beamforming matrix and $\boldsymbol{R}_0\triangleq\mathbb{E}\{\boldsymbol{x}_0\boldsymbol{x}_0^H\}$ denotes the BS sensing covariance matrix.

We consider a three-dimensional Cartesian coordinate system for the XL-RIS assisted ISAC system, where the XL-RIS is located in the $xoz$-plane, the BS is located in the $yoz$-plane, and both CUs and TGTs are located in the $xoy$-plane. First, we denote  
$\boldsymbol{c}^{\text{RIS}}_{n_x,n_z}=[c^{\text{RIS}}_{n_x},0, c^{\text{RIS}}_{n_z}]^T$ and $\boldsymbol{c}^{\text{BS}}_m=[c_{m,x}^{\text{BS}}, c_{m,y}^{\text{BS}}, c_{m,z}^{\text{BS}}]^T$ as the coordinates of the XL-RIS's reflecting element $(n_x,n_z)$ and the BS's antenna $m$, respectively, where $n_x\in{\cal N}_x\triangleq\{-\tilde{N}_x,-\tilde{N_x}+1,...,\tilde{N}_x\}$, $n_z\in{\cal N}_x\triangleq\{-\tilde{N}_z, -\tilde{N}_z+1,...,\tilde{N}_z\}$, and $m\in{\cal M}\triangleq\{1,2,...,M\}$. Denote by $\boldsymbol{c}^{\text{CU}}_k=[c_k^{\text{CU},x},c_k^{\text{CU},y},0]^T$ and $\boldsymbol{c}_l^{\text{TGT}}=[c_l^{\text{TGT},x},c_l^{\text{TGT},y},0]^T$ the coordinates of CU $k\in{\cal K}$ and TGT $l\in{\cal L}$, respectively.

As such, the Euclidean distances from the BS's antenna $m$ to the XL-RIS's element $(n_x,n_z)$ and CU $k\in \cal K$ are respectively given by
\begin{subequations}
 \begin{align}
       & r^{\text{BS-CU}}_{m,k} =\left\|\boldsymbol{c}^{\text{BS}}_m -\boldsymbol{c}^{\text{CU}}_{k}\right\|\label{eq.p5}\\
     & r^{\text{BS-R}}_{m,n_x,n_z} =\left\|\boldsymbol{c}^{\text{BS}}_m-\boldsymbol{c}^{\text{RIS}}_{n_x,n_z}\right\|, \label{eq.p6} 
 \end{align} 
\end{subequations}
where $m\in{\cal M}$, and the Euclidean distances from the XL-RIS's element $(n_x,n_z)$ to CU $k\in{\cal K}$ and TGT $l\in{\cal L}$ are respectively given by
\begin{subequations}
\begin{align}
   & r^{\text{R-CU}}_{n_x,n_z,k} =\left\|\boldsymbol{c}^{\text{RIS}}_{n_x,n_z} - \boldsymbol{c}^{\text{CU}}_{k}\right\|\label{eq.p31}\\
    & r^{\text{R-TGT}}_{n_x,n_z,l} =\left\| \boldsymbol{c}^{\text{RIS}}_{n_x,n_z} - \boldsymbol{c}^{\text{TGT}}_{l}\right\|. \label{eq.p3}
\end{align}
\end{subequations}
 
\subsection{XL-RIS assisted Communication Model}
Let $\boldsymbol{G}\in \mathbb{C}^{N\times{M}}$ denote the near-field channel matrix from the BS to the XL-RIS, which is given by
 \begin{equation}\label{eq.p10}
    \boldsymbol{G} = \beta^{\text{BS-R}}\begin{bmatrix}
  e^{-j \frac{2 \pi}{\lambda} r^{\text{BS-R}}_{1,-\tilde{N}_x,-\tilde{N}_z}}& \dots &e^{-j \frac{2 \pi}{\lambda} r^{\text{BS-R}}_{M,-\tilde{N}_x,-\tilde{N}_z}} \\
  \vdots &  &\vdots \\
  e^{-j \frac{2 \pi}{\lambda} r^{\text{BS-R}}_{1,0,0}}&   \dots&e^{-j \frac{2 \pi}{\lambda} r^{\text{BS-R}}_{M,0,0}} \\
  \vdots&   &\vdots \\
  e^{-j \frac{2 \pi}{\lambda} r^{\text{BS-R}}_{1,\tilde{N}_x,\tilde{N}_z}}&  \dots &e^{-j \frac{2 \pi}{\lambda} r^{\text{BS-R}}_{M,\tilde{N}_x,\tilde{N}_z}}
\end{bmatrix},
\end{equation}
where $\beta^{\text{BS-R}}=\rho_0(r^{\text{BS-R}}_{1,0,0})^{-\alpha}$ denotes the path-loss coefficient from the BS to XL-RIS. Furthermore, Denote by $\rho_0$ and $\alpha$ the path-loss coefficient of signal propagation at the reference distance of one meter and path-loss exponent, respectively.

Let $\psi^{\text{R-CU}}_{k}$ and $\phi^{\text{R-CU}}_{k}$ denote the azimuth and elevation angles of departure (AoDs) from the XL-RIS central element $(0,0)$ to CU $k\in{\cal K}$, respectively. Accordingly, the XL-RIS's array response with the AoD-distance-pair $(\psi^{\text{R-CU}}_{k},\phi^{\text{R-CU}}_{k},r^{\text{R-CU}}_{0,0,k})$ is given by\cite{b111}
\begin{align}
    \boldsymbol{\alpha}^{\text{R-CU}}_k(\psi^{\text{R-CU}}_{k},\phi^{\text{R-CU}}_{k},r^{\text{R-CU}}_{0,0,k}) = &
    \boldsymbol{\alpha}^{\text{R-CU}}_{k,x}\otimes \boldsymbol{\alpha}^{\text{R-CU}}_{k,z},  
\end{align}
where for each $n_x\in{\cal N}_x$ and $n_z\in{\cal N}_z$, it follows that
\begin{subequations}
\begin{align}
&[\boldsymbol{\alpha}_{k,x}^{\text{R-CU}}]_{n_x}= \exp\Bigg(-j \frac{2 \pi}{\lambda} \Big(-n_xd\cos\psi^{\text{R-CU}}_{k}\sin\phi^{\text{R-CU}}_{k} \notag \\
& \quad\quad\quad\quad\quad  +\frac{n_x^2d^2(1-\cos^2\psi^{\text{R-CU}}_{k}\sin^2\phi^{\text{R-CU}}_{k})}{2r^{\text{R-CU}}_{0,0,k}} \Big) \Bigg)\\
&[\boldsymbol{\alpha}_{k,z}^{\text{R-CU}}]_{n_z}=\exp\Bigg(-j \frac{2 \pi}{\lambda}\Big(
-n_zd\cos\phi^{\text{R-CU}}_{k} \notag\\
&\quad\quad\quad\quad\quad  +\frac{n_z^2d^2\sin^2\phi^{\text{R-CU}}_{k}}{2r^{\text{R-CU}}_{0,0,k}}
\Big) \Bigg).
\end{align}
\end{subequations}

Let $\beta^{\text{R-CU}}_{k}=\rho_0(r^{\text{R-CU}}_{0,0,k})^{-\alpha}$ denote the path-loss coefficient from the XL-RIS's element $(n_x,n_z)$ to CU $k$. The path-loss coefficient from the BS's antenna $m$ to CU $k$ is denoted as $\beta^{\text{BS-CU}}_{k}=\rho_0(r^{\text{BS-CU}}_{1,k})^{-\alpha}$. The channel vectors from the XL-RIS and BS to CU $k\in{\cal K}$ are then respectively modeled as
 \begin{subequations}
 \begin{align}
 &\boldsymbol{h}^{\text{R-CU}}_k=\beta^{\text{R-CU}}_{k}\boldsymbol{\alpha}^{\text{R-CU}}_k(\psi^{\text{R-CU}}_{k},\phi^{\text{R-CU}}_{k},r^{\text{R-CU}}_{0,0,k})\\
 &\boldsymbol{h}^{\text{BS-CU}}_{k} =\beta^{\text{BS-CU}}_{k} \boldsymbol{\alpha}_k^{\text{BS-CU}}, \label{eq.p7} 
 \end{align}
 \end{subequations}
where $\boldsymbol{\alpha}^{\text{BS-CU}}_k=[e^{-j \frac{2 \pi}{\lambda} r^{\text{BS-CU}}_{1,k}}, e^{-j \frac{2 \pi}{\lambda} r^{\text{BS-CU}}_{2,k}}, ...,e^{-j \frac{2 \pi}{\lambda} r^{\text{BS-CU}}_{M,k}}]^T$ denotes the array steering vector from the BS to CU $k\in{\cal K}$.

 %Then, the channel coefficient from the BS's antenna $m$ to CU $k\in{\cal K}$ is given by\cite{b10}
 %\begin{align}
 %h^{\text{BS-CU}}_{m,k} =\beta^{\text{BS-CU}}_{k} \exp(-j \frac{2 \pi}{\lambda} r^{\text{BS-CU}}_{m,k}), \label{eq.p7} 
 %\end{align}
 %where $m=1,2,...,M$.

% Based on \eqref{eq.p7}, the complex-valued near-field channel coefficient vector $\boldsymbol{h}_{bu,k}\in \mathbb{C}^{M\times{1}}$ from the BS and CU $k$ is given by
 %\begin{equation}\label{eq.p9}
  %  \boldsymbol{h}_{bu,k} = \left[h^{bu}_{-\tilde{M},k}, \ldots, h^{bu}_{\tilde{M},k}\right]^{T}.
%\end{equation}

 Let $\boldsymbol{\Phi}$ denote the XL-RIS reflecting matrix, where $\Phi_{n_x,n_z}=e^{j\phi_{n_x,n_z}}$ denotes the reflecting coefficient of the XL-RIS's element $(n_x,n_z)$ and $\phi_{n_x,n_z}\in[0,2\pi]$ denotes the phase shift value of the XL-RIS's element $(n_x,n_z)$, $\forall n_x\in{\cal N}_x$, $n_z\in{\cal N}_z$. For notational convenience, we define $\boldsymbol{\theta}=\text{vec}({\boldsymbol{\Phi}})$ as the XL-RIS reflecting vector which stacks the columns of $\boldsymbol{\Phi}$.

% $\boldsymbol{\Phi}=[ e^{j\phi_{-N_x,-N_z}},..., e^{j\phi_{-N_x+1,-N_z}},...,e^{j\phi_{\tilde{N_x},\tilde{N_z}}}]^{T}$ denote the phase shift vector of the RIS, where $\theta_{n_y,n_z}\in\Theta\triangleq\{e^{j \phi_{n_y,n_z}} \mid \phi_{n_y,n_z} \in(0,2 \pi]\}$, $\forall n_y\in{\cal N}_y$, $n_z\in{\cal N}_z$. 

In the ISAC system under consideration, each CU $k\in{\cal K}$ receives signals from the BS through a direct link and a cascaded link assisted by the XL-RIS. Therefore, the received signal of each CU $k\in{\cal K}$ is modeled as
 \begin{align}\label{eq.yk}
   y_{k} &= \big((\boldsymbol{h}^{\text{R-CU}}_k)^H \text{diag}(\boldsymbol{\theta}^H) \boldsymbol{G} + (\boldsymbol{h}^{\text{BS-CU}}_k)^{H}\big)\boldsymbol{x}+ n_k \notag \\
   &=(\boldsymbol{G}^H\text{diag}(\boldsymbol{h}^{\text{R-CU}}_k) \boldsymbol{\theta}   + \boldsymbol{h}^{\text{BS-CU}}_k)^H\boldsymbol{x}+ n_k \notag \\
   &= \boldsymbol{h}_{\text{CU},k}^H\boldsymbol{x} + n_k,
 \end{align}
 where $\boldsymbol{h}_{\text{CU},k} \triangleq \boldsymbol{G}^H\text{diag}(\boldsymbol{h}^{\text{R-CU}}_k)\boldsymbol{\theta}  + \boldsymbol{h}^{\text{BS-CU}}_{k}$ is referred to as the cascaded channel vector from BS to CU $k$ via the XL-RIS, and $n_k\sim \mathcal{CN}(0,\sigma_k^2)$ denotes the additive white Gaussian noise (AWGN) at CU $k$ with zero mean and variance $\sigma_k^2$. 
 
 Regarding decoding communication signals, the receiver of each CU $k\in{\cal K}$ treats all the sensing signal $\boldsymbol{x}_0$ and the communication signals for the other CUs as interference. Then, based on \eqref{eq.yk}, the decoding signal-to-interference-plus-noise ratio (SINR) of CU $k$ is expressed as
 \begin{equation}
    \gamma_{k} =  \frac{|\boldsymbol{h}_{\text{CU},k}^H \boldsymbol{w}_k|^2}{\sum_{j=1,j\neq k}^K|\boldsymbol{h}_{\text{CU},k}^H \boldsymbol{w}_j|^2 + \boldsymbol{h}_{\text{CU},k}^H\boldsymbol{R}_0\boldsymbol{h}_{\text{CU},k} + \sigma_k^2},
\end{equation}
where $k\in{\cal K}$.

\subsection{XL-RIS assisted Sensing Model}
 In the XL-RIS assisted ISAC system, the XL-RIS is enabled to flexibly reconfigure the BS $\to$ XL-RIS $\to$ TGT cascaded channel condition by adjusting the phase shift values of its reflecting elements, thereby providing an enhanced ISAC performance.
 
 Let $\psi^{\text{R-TGT}}_{l}$ and $\phi^{\text{R-TGT}}_{l}$ denote the azimuth and elevation AoDs from the XL-RIS central reflecting element $(0,0)$ to TGT $l\in{\cal L}$, respectively, and let $r^{\text{R-TGT}}_{0,0,l}$ denote the Euclidean distance from the XL-RIS central reflecting element $(0,0)$ to TGT $l\in{\cal L}$. For each TGT $l\in{\cal L}$ with the AoD-distance-pair $(\psi^{\text{R-TGT}}_{l,x},\phi^{\text{R-TGT}}_{l,x},r^{\text{R-TGT}}_{0,0,l})$, the near-field array response vector $\boldsymbol{\alpha}^{\text{R-TGT}}_{l}$ of the XL-RIS is given by
\begin{align}
    \boldsymbol{\alpha}^{\text{R-TGT}}_{l}(\psi^{\text{R-TGT}}_{l},\phi^{\text{R-TGT}}_{l},r^{\text{R-TGT}}_{0,0,l}) = \boldsymbol{\alpha}^{\text{R-TGT}}_{l,x}
    \otimes \boldsymbol{\alpha}^{\text{R-TGT}}_{l,z},
    \end{align}
where for each $n_x\in{\cal N}_x$ and $n_z\in{\cal N}_z$, it follows that
\begin{subequations}
\begin{align}
&[\boldsymbol{\alpha}_{l,x}^{\text{R-TGT}}]_{n_x}= \exp\Bigg(-j \frac{2 \pi}{\lambda} \Big(-n_xd\cos\psi^{\text{R-TGT}}_{l}\sin\phi^{\text{R-TGT}}_{l} \notag \\
& \quad\quad\quad  +\frac{n_x^2d^2(1-\cos^2\psi^{\text{R-TGT}}_{l}\sin^2\phi^{\text{R-TGT}}_{l})}{2r^{\text{R-TGT}}_{0,0,l}} \Big) \Bigg)\\
&[\boldsymbol{\alpha}_{l,z}^{\text{R-TGT}}]_{n_z}=\exp\Bigg(-j \frac{2 \pi}{\lambda}\Big(
-n_zd\cos\phi^{\text{R-TGT}}_{l} \notag\\
&\quad\quad\quad  +\frac{n_z^2d^2\sin^2\phi^{\text{R-TGT}}_{l}}{2r^{\text{R-TGT}}_{0,0,l}}
\Big) \Bigg).
\end{align}
\end{subequations}

Denote by $\beta_l^{\text{RIS-TGT}}$ the path-loss coefficient of signal propagation from the XL-RIS central element $(0,0)$ to TGT $l\in{\cal L}$. The channel vector from the XL-RIS to TGT $l\in{\cal L}$ is
 \begin{align}
\boldsymbol{h}^{\text{R-TGT}}_l=\beta^{\text{R-TGT}}_{l}\boldsymbol{\alpha}^{\text{R-TGT}}_l(\psi^{\text{R-TGT}}_{l},\phi^{\text{R-TGT}}_{l},r^{\text{R-TGT}}_{0,0,l}).
 \end{align}

Note that the communication signal $\sum_{k=1}^K\boldsymbol{w}_ks_k$ and sensing signal $\boldsymbol{x}_0$ in the BS transmit signal $\boldsymbol{x}$ can both be applied to illuminate a number of $L$ TGTs via the BS $\to$ XL-RIS $\to$ TGT cascaded channels. Therefore, the sensing beampattern gain $\rho_l$ for TGT $l\in{\cal L}$ is expressed as
      \begin{align}        
  \rho_l
       % &=\mathbb{E}[  |\boldsymbol{h}_{\text{TGT},l}^{H}\boldsymbol{x}|^2 ]\\
    % &=\boldsymbol{h}_{\text{TGT},l}^{H}\Big( \sum_{k=1}^K\boldsymbol{w}_k \boldsymbol{w}^H_k + \boldsymbol{R}_0\Big) \boldsymbol{h}_{\text{TGT},l}\\
  &=\boldsymbol{h}_{\text{TGT},l}^{H}\big(\boldsymbol{W} \boldsymbol{W}^H + \boldsymbol{R}_0\big) \boldsymbol{h}_{\text{TGT},l},
    \end{align}
 where $\boldsymbol{h}_{\text{TGT},l} \triangleq \boldsymbol{G}^H\text{diag}(\boldsymbol{h}^{\text{R-TGT}}_l) \boldsymbol{\theta}$ denotes the BS $\to$ XL-RIS $\to$ TGT cascaded channel vector for TGT $l\in{\cal L}$.
 
\subsection{Problem Formulation}
Subject to the minimal sensing beampattern gain requirement of all the $L$ TGTs and the BS transmit power constraint, we aim to maximize the WSR of all the $K$ CUs by jointly optimizing BS communication beamforming matrix $\boldsymbol{W}$ and sensing covariance matrix $\boldsymbol{R}_0$, as well as the XL-RIS's reflecting coefficient vector $\boldsymbol{\theta}$. Mathematically, the joint near-field ISAC beamforming design problem is described as
\begin{subequations}\label{eq.P1}
\begin{align}
  (\text{P}1):~ &\max_{\boldsymbol{W},\boldsymbol{R}_0,\boldsymbol{\theta}}~ \sum_{k=1}^K \tau_k\log \left(1+\gamma_k\right)\label{eq.p111}\\
  \text{s.t.}~~& \text{tr}(\boldsymbol{W}^H\boldsymbol{W})  + \text{tr}(\boldsymbol{R}_0)\leq P_0\label{eq.p11}\\
 &\left |  \Phi_{n_x,n_z}\right |=1,~ \forall n_x\in{\cal N}_x, n_z\in{\cal N}_z\label{eq.p12}\\
 & \rho_l\ge\rho_{th}\label{eq.p15d}, \forall l\in{\cal L},
\end{align}
\end{subequations}
where $\tau_k\in(0,1)$ denotes the priority weight of CU $k$ in downlink achievable data rate, $P_0$ denotes the BS transmit power budget, and $\rho_{th}$ denotes the sensing beampattern gain threshold for all the TGTs. 

% Our goal is to maximize communication performance \eqref{eq.p111} under reasonable power constraint \eqref{eq.p11}, while maintaining sensing performance of the ISAC system via the inequality constraints in \eqref{eq.p15d}.

Note that the BS transmit design variables $(\boldsymbol{W},\boldsymbol{R}_0)$ and the XL-RIS's reflecting coefficient vector $\boldsymbol{\theta}$ are strongly coupled. Hence, (P1) is a nonconvex optimization. As a compromise, we obtain a local optimal solution of (P1) by applying the classic BCD method in a low complexity fashion in Section III. In addition, in Section~IV we develop a state-of-the-art GNN based solution for (P1) by training a deep learning model with the graph-structured XL-RIS assisted near-field ISAC data, achieving superior performance with an affordable complexity. 

%Since the optimal solution is irrelevant to the base of the logarithm function, we use the natural logarithm throughout the paper.

\section{FP based BCD Solution for (P1)}\label{3}

In this section, we develop a FP based BCD algorithm for solving the joint near-field ISAC design problem (P1) .

\subsection{Closed-Form FP Reformulation}

Based on the idea of fractional program (FP)\cite{b16,b161}, i.e., $ \log(1+\gamma_k) = \max_{\lambda_k>0} ( \log(1+\lambda_k) - \lambda_k +\frac{(1+\lambda_k)\gamma_k}{1+\gamma_k} )$, the weighted rate maximization problem (P1) is recast as
\begin{align*}
&(\text{P1.1})~\max_{\boldsymbol{W},\boldsymbol{R}_0,\boldsymbol{\theta},\boldsymbol{\lambda}}~ \sum_{k=1}^K \tau_k\Big(\log (1+\lambda_k)-\lambda_k+\frac{(1+\lambda _k)\gamma_k}{1+\gamma_k}\Big)\\
&\text{s.t.}~~\eqref{eq.p11},\eqref{eq.p12},\eqref{eq.p15d}\\
&~~~~~~~\lambda_k\ge 0,\forall k\in{\cal K},
\end{align*}
where $\boldsymbol{\lambda}=[\lambda_1,...,\lambda_K]^T$. Note that given vector $\boldsymbol{\lambda}$, maximizing the objective function of problem (P1.1) is equivalent to maximizing the ratio function $\sum_{k=1}^K \frac{(1+\lambda_k)\gamma_k}{1+\gamma_k}$, where
\begin{align}\label{eq.ratio}
    \frac{(1+\lambda_k)\gamma_k}{1+\gamma_k} =  \frac{|\sqrt{1+\lambda_k}\boldsymbol{h}_{\text{CU},k}\boldsymbol{w}_k|^2}{\boldsymbol{h}_{\text{CU},k}(\boldsymbol{W}\boldsymbol{W}^H+\boldsymbol{R}_0)\boldsymbol{h}_{\text{CU},k}+\sigma_k^2}.
\end{align}

Furthermore, by introducing an auxiliary variable $v=\frac{A(x)}{B(x)}$, it follows that
\begin{align} \label{eq.beta}
  \max_x \frac{|A(x)|^2}{B(x)}=\max_{x,v} (v A^*(x) + v^*A(x) - |\beta|^2B(x)). 
\end{align} 

Therefore, based on \eqref{eq.ratio} and \eqref{eq.beta}, problem (P1.1) is equivalently reformulated as a biconvex problem \begin{subequations}\label{eq.P1.1}
\begin{align}
  (\text{P}1.2):~ &\min_{\boldsymbol{\lambda}, \boldsymbol{v}, \boldsymbol{W},\boldsymbol{R}_0,\boldsymbol{\theta}} ~-f(\boldsymbol{\lambda}, \boldsymbol{v}, \boldsymbol{W},\boldsymbol{R}_0,\boldsymbol{\theta}) \label{eq.P1.11}\\
  \text{s.t.}~~& \lambda_k\ge 0, \forall{k=1,...,K} \\   &\eqref{eq.p11},\eqref{eq.p12},\eqref{eq.p15d},
\end{align}
\end{subequations}
where $\boldsymbol{v}=[v_1,...,v_K]^T$, and the function $f$ is defined as
\begin{align}
    &f(\boldsymbol{\lambda}, \boldsymbol{v}, \boldsymbol{W},\boldsymbol{R}_0,\boldsymbol{\theta})=\sum_{k=1}^K\tau_k(\log(1+\lambda_k) -\lambda_k) \notag \\
    &+\sum_{k=1}^K \sqrt{\tau_k( 1+\lambda_k)}(v_k \boldsymbol{w}_k^H\boldsymbol{h}_{\text{CU},k}+v_k^* \boldsymbol{h}_{\text{CU},k}^H\boldsymbol{w}_k)    \notag \\
    &-\sum_{k=1}^K\left | v_k  \right |^2  \big( \boldsymbol{h}_{\text{CU},k}^H \big(\boldsymbol{W}\boldsymbol{W}^H+\boldsymbol{R}_0\big)\boldsymbol{h}_{\text{CU},k} +\sigma_k^2\big).
\end{align}  

Note that $f$ is a concave function in each variable block of $\boldsymbol{\lambda}$, $\boldsymbol{v}$, $(\boldsymbol{W},\boldsymbol{R}_0)$ and $\boldsymbol{\theta}_0$. Therefore, we next employ the BCD method to obtain a local-optimal solution of $(\text{P}1.2)$.

\subsection{Optimization of $(\boldsymbol{\lambda},\boldsymbol{v})$}

 %, where the variables of $\{\boldsymbol{\lambda}, \boldsymbol{v}, \boldsymbol{W},\boldsymbol{R}_0,\boldsymbol{\theta}\}$ are iteratively updated in an alternating manner. 

Denote by $\bar{\boldsymbol{\lambda}}, \bar{\boldsymbol{v}}, \bar{\boldsymbol{W}},\bar{\boldsymbol{R}}_0,\bar{\boldsymbol{\theta}}$ the temporal optimization results in the previous iteration. Based on \cite{b16}, the closed-form of $(\boldsymbol{\lambda},\boldsymbol{v})$ is given by
\begin{subequations}
\begin{align}
   &\lambda_k= \frac{\bar{\chi}_k^2 +\bar{\chi }_k\sqrt{\bar{\chi}_k^2+4} }{2} \label{eq.lambda_k}\\
    &v_k= \frac{\sqrt{\tau_k( 1+\bar{\lambda}_k  )}\bar{\boldsymbol{h}}^H_{\text{CU},k} \bar{\boldsymbol{w}}_k }{\bar{\boldsymbol{h}}_{\text{CU},k}^H(\bar{\boldsymbol{W}}\bar{\boldsymbol{W}}^H + \bar{\boldsymbol{R}_0})\bar{\boldsymbol{h}}_{\text{CU},k}+\sigma_k^2}, \label{eq.v_k}
\end{align}
\end{subequations}
where $\bar{\chi }_k=\frac{1}{\sqrt{\tau_k}}\Re\left \{ \bar{v}_k^* \bar{\boldsymbol{h}}_{\text{CU},k}^H \bar{\boldsymbol{w}}_k\right \}$, and $\bar{\boldsymbol{h}}_{\text{CU},k}=\boldsymbol{G}^H\text{diag}(\boldsymbol{h}^{\text{R-CU}}_k) \bar{\boldsymbol{\theta}}  + \boldsymbol{h}^{\text{BS-CU}}_k$ denotes the effective channel from the BS to CU $k$ in the previous iteration.

\subsection{Optimization of Transmission Design $(\boldsymbol{W},\boldsymbol{R}_0)$}

Given $\bar{\boldsymbol{\lambda}}, \bar{\boldsymbol{v}},\bar{\boldsymbol{\theta}}$, we next optimize the BS transmit design variables $(\boldsymbol{W},\boldsymbol{R}_0)$ for $K$ CUs and $L$ TGTs. Specifically, by removing the constant terms independent of $\boldsymbol{W}$ and $\boldsymbol{R}_0$ in $f(\boldsymbol{\lambda},\boldsymbol{v},\boldsymbol{W},\boldsymbol{R}_0,\boldsymbol{\theta})$, the optimization of $(\boldsymbol{W},\boldsymbol{R}_0)$ is expressed as the following minimization problem
\begin{subequations}\label{eq.WR0}
    \begin{align}
&(\boldsymbol{\hat W},\boldsymbol{\hat R}_0) = \arg\min_{\boldsymbol{W},\boldsymbol{R}_0}
g(\boldsymbol{W},\boldsymbol{R}_0 )\\
    &\text{s.t.}~\text{tr}(\boldsymbol{W}^H\boldsymbol{W}) + \text{tr}(\boldsymbol{R}_0)\leq P_0\label{eq.power}\\
    &~~~~\boldsymbol{\bar h}^H_{\text{TGT},l}(\boldsymbol{W}\boldsymbol{W}^H+\boldsymbol{R}_0)\boldsymbol{\bar h}_{\text{TGT},l} \ge\rho_{th},\forall l\in{\cal L}\label{eq.sub1},
\end{align}\label{eq.sub0}
\end{subequations}
where $\boldsymbol{\bar h}_{\text{TGT},l} \triangleq \boldsymbol{G}^H\text{diag}(\boldsymbol{h}^{\text{R-TGT}}_l) \boldsymbol{\bar \theta}$, and the function $g$ is  
\begin{align}
g(\boldsymbol{W},\boldsymbol{R}_0)=
    &\sum_{k=1}^K|\bar{v}_k|^2    \bar{\boldsymbol{h}}_{\text{CU},k}^H(\boldsymbol{W}\boldsymbol{W}^H+\boldsymbol{R}_0)\bar{\boldsymbol{h}}_{\text{CU},k}   \nonumber \\
    &-\sum_{k=1}^K 2\sqrt{\tau_k( 1+\bar{\lambda}_k)}\Re\{ \bar{v} _k^* \bar{\boldsymbol{h}}_{\text{CU},k}^H \boldsymbol{w}_k \}.
\end{align}

Note that due to the non-convexity of constraints in \eqref{eq.sub1}, problem \eqref{eq.WR0} is a non-convex optimization problem. As such, we employ the celebrated SCA technique to obtain a first-order linearization for the constraint \eqref{eq.sub1}, which is given as
\begin{align}\label{eq.sca_W}
  2\Re\{  \boldsymbol{h}_{\text{TGT},l}^H{\boldsymbol{W}}{\boldsymbol{\bar W}}^H\boldsymbol{\bar h}_{\text{TGT},l}\} + \boldsymbol{\bar h}_{\text{TGT},l}^H\boldsymbol{R}_0\boldsymbol{\bar h}_{\text{TGT},l}^H \notag \\
    \ge \rho_{\text{th}}+\boldsymbol{\bar h}_{\text{TGT},l}^H{\boldsymbol{\bar W}}{\boldsymbol{\bar W}}^H\boldsymbol{\bar h}_{\text{TGT},l},\forall l\in{\cal L}.  
\end{align}

In this way, problem \eqref{eq.WR0} can be approximated as a convex quadratically constrained quadratic program (QCQP), i.e., 
\begin{subequations}\label{eq.WR0_Convex}
\begin{align}
&\min_{\boldsymbol{W},\boldsymbol{R}_0} g(\boldsymbol{W},\boldsymbol{R}_0)+g_{\text{prox1}}(\boldsymbol{W})+g_{\text{prox2}}(\boldsymbol{R}_0)\\
&\text{s.t.}~\eqref{eq.power},\eqref{eq.sca_W},
\end{align}  
\end{subequations}
where $g_{\text{prox1}}(\boldsymbol{W})=\frac{C_1}{2}\sum_{k=1}^K  \left \| \boldsymbol{w}_k-\overline{\boldsymbol{w}}_k \right \|^2 $ and $g_{\text{prox2}}(\boldsymbol{R}_0)=\frac{C_2}{2}\left \| \boldsymbol{R}_0-\overline{\boldsymbol{R}}_0 \right \|_F^2$ denote the proximal regularization terms corresponding to variables $\boldsymbol{W}$ and $\boldsymbol{R}_0$, respectively \cite{b162}, $C_1>0$ and $C_2>0$ denote the proximal regularization parameters. Note that these quadratic terms $g_{\text{prox1}}$ and $g_{\text{prox2}}$ are responsible for controlling the variable update step size to prevent oscillation. Problem \eqref{eq.WR0_Convex} is a strictly convex problem. Therefore, the optimal solution $(\boldsymbol{\hat W}$ and $\boldsymbol{\hat R}_0)$ for \eqref{eq.WR0_Convex} can be efficiently obtained  by using standard convex optimization solvers, such as the CVX toolbox \cite{b15}.

\subsection{Optimization of XL-RIS Reflecting Coefficient Vector $\boldsymbol{\theta}$}
Now given $\bar{\boldsymbol{\lambda}}, \bar{\boldsymbol{v}}, \bar{\boldsymbol{W}},\bar{\boldsymbol{R}}_0$, we optimize the XL-RIS reflecting coefficient vector $\boldsymbol{\theta}$. For notational convenience, we define $\boldsymbol{H}_{RT,l}=\text{diag}((\boldsymbol{h}^{\text{R-TGT}}_l)^H)\boldsymbol{G}\in\mathbb{C}^{N\times M}$, $\forall l\in{\cal L}$ and $\boldsymbol{H}_{RC,k}=\text{diag}((\boldsymbol{h}^{\text{R-CU}}_k)^H)\boldsymbol{G}\in\mathbb{C}^{N\times M}$, $\forall k\in{\cal K}$. By dropping irrelevant constant terms of $\boldsymbol{\theta}$ in $f(\boldsymbol{\lambda},\boldsymbol{v},\boldsymbol{W},\boldsymbol{R}_0,\boldsymbol{\theta})$, the optimization of $\boldsymbol{\theta}$ is formulated as the following minimization problem as 
\begin{subequations}\label{eq.theta_opt}
   \begin{align}
    &\hat{\boldsymbol{\theta}}=\arg \min_{\boldsymbol{\theta}}
  \boldsymbol{\theta}^H\boldsymbol{B}\boldsymbol{\theta}- 2\Re\{  \boldsymbol{\theta}^H \boldsymbol{\eta} \}\\
&\text{s.t.}~\left |  \Phi_{n_x,n_z}\right |=1,\forall n_x\in{\cal N}_x,n_z\in{\cal N}_z\label{module1} \\
    &~~~~\boldsymbol{\theta}^H\boldsymbol{A}_l\boldsymbol{\theta} \ge\rho_{th},\forall l\in{\cal L}\label{theta21}, 
\end{align} 
\end{subequations}
where $\boldsymbol{A}_l\succeq 0$, $\boldsymbol{B}\succeq 0$, and $\boldsymbol{\eta}$ are respectively given as  
\begin{align*}
&\boldsymbol{A}_l=\boldsymbol{H}_{RT,l}\boldsymbol{Q}\boldsymbol{H}^H_{RT,l},\forall l\in{\cal L}\\
   &   \boldsymbol{B}=\sum_{k=1}^K |\bar{v}_k|^2  \boldsymbol{H}_{RC,k}\boldsymbol{Q}\boldsymbol{H}_{RC,k}^H \\
    &\boldsymbol{\eta}=\sum_{k=1}^K  \big(\sqrt{\tau_k( 1+\bar{\lambda}_k)} \bar{v}_k^* 
    \boldsymbol{H}_{RC,k}\boldsymbol{\bar w}_k - 
  |\bar{v}_k|^2 \boldsymbol{H}_{RC,k}\boldsymbol{Q}\boldsymbol{h}^{\text{R-CU}}_k\big)
\end{align*}
with $\boldsymbol{Q}=
   \boldsymbol{\bar W}\boldsymbol{\bar W}^H+\boldsymbol{\bar R}_0$.
 
Note that the unit modulus constraints on the XL-RIS phase shift in \eqref{module1} define a non-convex set for $\boldsymbol{\theta}$, and the superlevel set of each convex quadratic function in the left hand side of \eqref{theta21} is not a convex set. Therefore, problem \eqref{eq.theta_opt} is a non-convex optimization problem, which is challenging to efficiently obtain its global solution.  

Nonetheless, due to the unit modulus constraints $\left |  \Phi_{n_x,n_z}\right |=1$, $\forall n_x\in{\cal N}_x, n_z\in{\cal N}_z$, it is shown that the feasible variable $\boldsymbol{\theta}$ of \eqref{eq.theta_opt} lies on a Riemannian manifold. Next, we employ a {\em penalty-based Riemannian manifold optimization} algorithm \cite{b163} to efficiently solve \eqref{eq.theta_opt}. 

Let $\mathcal{M}_{R}$ denote the Riemannian manifold which contains the feasible $\boldsymbol{\theta}$ of problem \eqref{eq.theta_opt}. It then follows that
\begin{equation}
   \mathcal{M}_R \triangleq \{ \boldsymbol{\theta} \in \mathbb{C}^N : \boldsymbol{\theta} \odot \boldsymbol{\theta}^*=\mathbf{1}_N  \},
\end{equation}
where $\odot$ denotes Hadamard product. Based on the Riemannian manifold optimization framework, the sensing beampattern gain constraints \eqref{theta21} are transformed into a quadratic penalty term in the objective function. Therefore, problem \eqref{eq.theta_opt} is reformulated as 
\begin{align}\label{eq.theta_opt2}
    \boldsymbol{\hat \theta}=\arg \min_{\boldsymbol{\theta} \in \mathcal{M}_R}
  f_R(\boldsymbol{\theta}),
\end{align}
where the objective function is given by
\begin{equation}
    f_R(\boldsymbol{\theta})=\boldsymbol{\theta}^H\boldsymbol{B}\boldsymbol{\theta}- 2\Re\{  \boldsymbol{\theta}^H \boldsymbol{\eta}  \}+
    f_{\text{pen}}(\boldsymbol{\theta})\label{penalty}
\end{equation}
with $f_{\text{pen}}(\boldsymbol{\theta})=    \frac{C}{2}\sum_{l=1}^L ( (\rho_{th}-\boldsymbol{\theta}^H\boldsymbol{A}_l\boldsymbol{\theta} )^+ ) ^2$ being the quadratic penalty term, and $C>0$ denotes the prescribed penalty factor.

For problem \eqref{eq.theta_opt2}, we employ the Riemannian conjugate gradient (RCG) algorithm \cite{b18,b19,b20}, in which the geometric properties of manifold ${\cal M}_R$ are leveraged to iteratively update the XL-RIS reflecting coefficient vector $\boldsymbol{\theta}$ via the gradient descent on the tangent space and retraction operations. In particular, the RCG algorithm consists of the following three key steps in each iteration.
\begin{itemize}
    \item \emph{Step 1-Obtain Riemannian Gradient:} The Riemannian gradient of $f_R(\boldsymbol{\theta})$ with respect to $\boldsymbol{\theta}$, denoted by $\text{grad} f_R(\boldsymbol{\theta})$, is defined as the orthogonal projection of the Euclidean gradient $\nabla f_R(\boldsymbol{\theta})$ onto the complex circle. Then, the Riemannian gradient $\text{grad} f_R(\boldsymbol{\theta})$ during each iteration is computed by
\begin{align}
    \text{grad}f_R(\boldsymbol{\theta})=\nabla f_R(\boldsymbol{\theta})-\Re\{ \nabla f_R(\boldsymbol{\theta}) \odot \boldsymbol{\theta}^*\} \odot \boldsymbol{\theta},
\end{align}
where the Euclidean gradient is given by
$\nabla f_R(\boldsymbol{\theta}) =\boldsymbol{B}\boldsymbol{\theta}-\boldsymbol{\eta}-C\sum_{l=1}^L  (\rho_{th}-\boldsymbol{\theta}^H\boldsymbol{A}_l\boldsymbol{\theta} )^+\boldsymbol{A}_l\boldsymbol{\theta}$.

\item \emph{Step 2-Find Search Direction:} Denote by $\boldsymbol{d}$ and $\bar{\boldsymbol{d}}$ the search directions at the current and previous iterations, respectively. During each iteration, the tangent vector conjugate to $\text{grad}f_{\text{R}}(\boldsymbol{\theta})$ is determined as the search direction $\boldsymbol{d}$. Correspondingly, it follows that 
\begin{align}
    \boldsymbol{d}=-\text{grad}f_R(\boldsymbol{\theta})+C_0(\boldsymbol{\bar d}-\Re\{ \boldsymbol{d} \odot \boldsymbol{\theta}^* \} \odot \boldsymbol{\theta}),
\end{align}
where $C_0$ denotes the gradient update parameter.

\item \emph{Step 3-Retraction:} Note that the tangent vector needs to be projected back to the complex circle manifold. Correspondingly, in one iteration, the XL-RIS reflecting coefficient vector $\boldsymbol{\theta}$ is updated as
\begin{align}
[\boldsymbol{\theta}]_n=\frac{[\boldsymbol{\theta}]_n+p[\boldsymbol{d}]_n }{| [\boldsymbol{\theta}]_n+p[\boldsymbol{d}]_n|},~n=1,...,N,
\end{align}
where $p>0$ denotes the Armijo-Goldstein step size. 

% Table~\ref{tab:mapping} illustrate the relationship between $n$ and the pair $(n_x,n_z)$.
\end{itemize}

% \begin{table}[!]
%   \centering
%   \caption{Mapping Relationship between $n$ and $(n_x, n_y)$}
%   \label{tab:mapping}
%   %\renewcommand{\arraystretch}{1.2}
%   \begin{tabular}{ccc}
% \toprule
%     \textbf{Linear Index} $n$ & \textbf{$x$-axis Index} $n_x$ & \textbf{$z$-axis Index} $n_y$ \\
%     \midrule
%     1 & $-\tilde{N}_x$ & $-\tilde{N}_z$ \\
%     2 & $-\tilde{N}_x$ & $-\tilde{N}_z+1$ \\
%     3 & $-\tilde{N}_x$ & $-\tilde{N}_z+2$ \\
%     $\vdots$ & $\vdots$ & $\vdots$ \\
%     $N_z$ & $-\tilde{N}_x$ & $\tilde{N}_z$ \\
%     $N_z + 1$ & $-\tilde{N}_x+1$ & $-\tilde{N}_z$ \\
%     $\vdots$ & $\vdots$ & $\vdots$ \\
%     $N = N_x \times N_z$ & $N_x$ & $N_z$ \\
%     \bottomrule
%   \end{tabular}
% \end{table}

\begin{algorithm}[h]
\caption{Proposed BCD-based Algorithm for (P1)}
\label{alg:BCD}
\begin{algorithmic}[1]
\STATE \textbf{Initialize}: Initialize $(\boldsymbol{W}^{(0)},\boldsymbol{R}_0^{(0)},\boldsymbol{\theta}^{(0)})$ that satisfy constraints and the auxiliary variables $(\boldsymbol{\lambda}^{(0)}, \boldsymbol{v}^{(0)}$), and set iteration index $t=0$.
\REPEAT
    \STATE Set $t \leftarrow t + 1$.
    \STATE \textit{Step 1: Update Auxiliary Variables $\boldsymbol{\lambda}$ and $\boldsymbol{v}$}
    \STATE Update $\boldsymbol{\lambda}^{(t)}$ and $\boldsymbol{v}^{(t)}$ according to the closed-form solutions in \eqref{eq.lambda_k} and \eqref{eq.v_k}, respectively.
    
    \STATE \textit{Step 2: Update XL-RIS Coefficients $\boldsymbol{\theta}$}
    \STATE Construct the penalized objective function $f_{\text{R}}(\boldsymbol{\theta})$.
    \STATE Update $\boldsymbol{\theta}^{(t)}$ via the RCG algorithm.
    
    \STATE \textit{Step 3: Update Transmit Beamforming $(\boldsymbol{W},\boldsymbol{R}_0)$}
    \STATE Construct the convex sub-problem for $(\boldsymbol{W},\boldsymbol{R}_0)$.
    \STATE Update $\boldsymbol{W}^{(t)}$ and $\boldsymbol{R}_0^{(t)}$ by solving \eqref{eq.WR0_Convex} based on the CVX toolbox.
\UNTIL The objective value between two iterations is smaller than the tolerance level $\epsilon$ or the maximum number of iterations $I_{max}$ is reached.
\end{algorithmic}
\end{algorithm}

\subsection{Proposed Algorithm and Complexity Analysis}
Algorithm~1 summarizes the complete proposed BCD procedure. Denote by $\epsilon>0$ the error tolerance level. The optimization variables are updated cyclically in the order of  $
    \cdots \boldsymbol{\lambda}, \boldsymbol{v}, \boldsymbol{\theta}, \boldsymbol{v}, (\boldsymbol{W},\boldsymbol{R}_0), \boldsymbol{\lambda} \cdots$
until the objective function converges. The computation complexity of the proposed BCD is discussed as follows.
\begin{itemize}
    \item \emph{Updating $(\boldsymbol{\lambda},\boldsymbol{v})$}: The closed-form solutions in \eqref{eq.lambda_k} and \eqref{eq.v_k} for updating $\boldsymbol{\lambda}$ and $\boldsymbol{v}$ are respectively presented, and the computational complexity is about $\mathcal{O}(KNM)$.
\item \emph{Updating $(\boldsymbol{W},\boldsymbol{R}_0)$}: The problem \eqref{eq.WR0_Convex} of optimizing $(\boldsymbol{W},\boldsymbol{R}_0)$ is a convex QCQP. If it is solved by standard convex solvers such as CVX toolbox, the complexity is approximately $\mathcal{O}(KM^3)$.
\item \emph{Updating $\boldsymbol{\theta}$}: The computational complexity of updating $\boldsymbol{\theta}$ is mainly determined by the RCG algorithm, which is about $\mathcal{O}(I_{RCG}(K^2N^2+LN^2))$ with $I_{RCG}$ being the number of iterations required for the RCG algorithm.
\end{itemize}

Consequently, the overall complexity of the proposed BCD method for (P1) is about $\mathcal{O}(I_{\text{out}}(KNM+KM^2+I_{RCG}(K^2N^2+LN^2)))$, where $I_{\text{out}}$ denotes the number of iterations.

\section{GNN-Based Solution for (P1)}\label{4}

In this section, we introduce the GNN model for characterizing the proposed XL-RIS assisted near-field ISAC system, and then propose a GNN based solution for (P1) that generates joint ISAC beamforming strategies by exchanging and updating feature information embedded in the GNN model.

\subsection{Basics of GNN}
As a class of DL models based on an information diffusion mechanism, the GNN model is able to extract key features from graph-based data effectively. The undirected graph is composed of a GNN-unit set and an edge set, where each unit corresponds to a node\cite{b201}. The output of a GNN is generated based on the states of units. Specifically, GNN follows a neighborhood aggregation scheme, where the node feature vector is computed by aggregating and concatenating features of its neighboring nodes\cite{b21}. In general, the architecture of a GNN is composed of three functional modules: an initial layer, multiple cascaded node update layers, and a readout layer.

Consider a graph $\mathcal{G} = (\mathcal{V}_{\text{g}}, \mathcal{E}_{\text{g}})$, where $\mathcal{V}_{\text{g}}$ and $\mathcal{E}_g$ the set of nodes (corresponding to XL-RIS, CUs, and TGTs) and the set of edges characterizing their interactions (e.g., signal/interference channel links), respectively. Generally, the node $v \in \mathcal{V}_{\text{g}}$ is initialized with a feature $\boldsymbol{u}_{v}^{(0)}$. Following a message-passing strategy, the node representation is iteratively updated by aggregating the representations of its neighbors. After $d_{\text{g}}$ rounds of message passing, the receptive field of each node is expected to expand to its $d_{\text{g}}$-hop neighborhood. Denote by $\boldsymbol{u}_{{\text{g}},v}^{(d_{\text{g}})}$ the feature of node $v$ at the $d_{\text{g}}$-th layer of GNN. The feature vector update rule is given by 
\begin{equation}
\boldsymbol{u}_{{\text{g}},v}^{(d_{\text{g}})} = f_{\text{comb}}^{(d_{\text{g}})} \left( \boldsymbol{u}_{{\text{g}},v}^{(d_{\text{g}}-1)}, f_{\text{agg}}^{(d_{\text{g}})} \left( \{ \boldsymbol{u}_{\text{g},v'}^{(d_{\text{g}}-1)}\}_{v' \in \mathcal{N}(v)} \right) \right),
\end{equation}
where $\mathcal{N}(v)$ denotes the set of neighboring nodes of $v$, and $f^{(d_g)}_{\text{agg}}$ and $f^{(d_g)}_{\text{comb}}$ denote the aggregate and combine functions at the $d_g$-th round of message passing, respectively. For the sake of brevity, we refer to the XL-RIS node simply as the RIS node in the following description.

\begin{figure}
    \centering
    \includegraphics[width=3.5in]{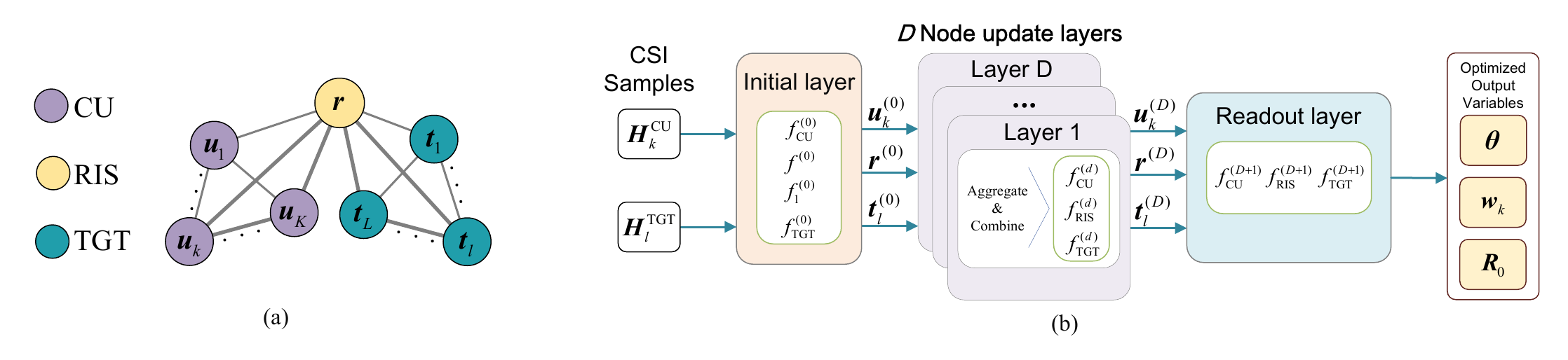}
    \caption{(a) Graph model of the XL-RIS assisted near-field ISAC systems; (b) Proposed GNN structure.}\label{fig_all_graph}
\end{figure}

\subsection{Graph Model of XL-RIS assisted Near-Field ISAC}

Fig.~\ref{fig_all_graph}(a) shows the established unweighted undirected graph model for the XL-RIS assisted ISAC system under consideration (including one RIS node, $K$ CU nodes and $L$ TGT nodes). Denote by  $\boldsymbol{r}$, $\boldsymbol{u}_k$, and $\boldsymbol{t}_l$ the node feature vectors of the RIS, CU $k$, and TGT $l$, respectively. Note that the RIS node feature vector $\boldsymbol{r}$ is associated with $K$ CU nodes and $L$ TGT nodes to acquire entire information for providing communication and sensing service, and there exists no edge connecting CU nodes to TGT nodes. Due to the presence of interference, there exists an edge connecting each other within the CU nodes and the TGT nodes. As a result, there are a total number of $K+L+\binom{K}{2}+\binom{L}{2}=\frac{K(K+1)+L(L+1)}{2}$ edges in the graph representation.

\subsection{Proposed GNN Structure}

In this subsection, we propose a GNN structure for the XL-RIS assisted ISAC system as shown in Fig.~\ref{fig_all_graph}(b), which consists of one initial layer, a number of $D$ update layers and one readout layer based on the graph model of the near-field ISAC. The BS transmit beamforming variables $({\{\boldsymbol{w}_k\}}_{k=1}^K,\boldsymbol{R}_0)$ and the XL-RIS reflecting coefficient vector $\boldsymbol{\theta}$ are both regarded as the output of the readout layer of the proposed GNN structure.

\subsubsection{Initial Layer}
\begin{figure}[H]
    \centering
    \includegraphics[width=2.2in]{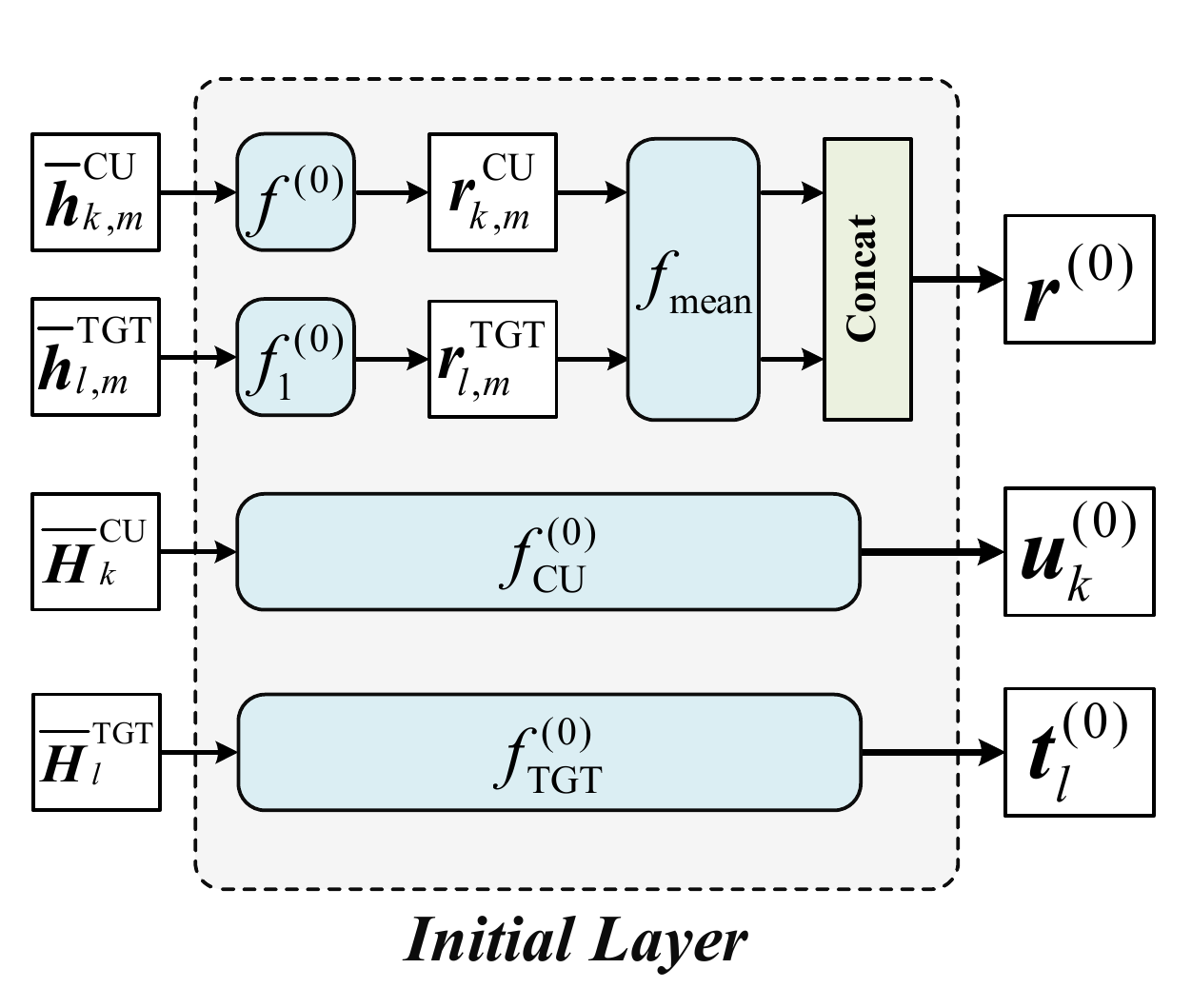}
    \caption{The structure of initial layer.}\label{figinit}
\end{figure}

Fig.~\ref{figinit} shows the structure of initial layer in the proposed GNN. Denote by $\boldsymbol{r}^{(0)}$, $\boldsymbol{u}_k^{(0)}$ and $\boldsymbol{t}_l^{(0)}$ the initial node feature vectors of the BS, CU $k\in{\cal K}$, and TGT $l\in{\cal L}$, respectively. For the initial layer, the input is the CSI from the BS to CUs/TGTs with the XL-RIS assistance for ISAC, i.e., 
\begin{subequations}
    \begin{align}  
    & \boldsymbol{H}_k^{\text{CU}}\triangleq\left[\boldsymbol{G}^H\text{diag}(\boldsymbol{h}^{\text{R-CU}}_k),\boldsymbol{h}^{\text{BS-CU}}_k\right] \in\mathbb{C}^{M  \times (N+1)}\\
    & \boldsymbol{h}_{k,m}^{\text{CU}}=(\boldsymbol{H}_k^{\text{CU}}(m,:))^T\in\mathbb{C}^{(N+1)  \times 1}\\
&\boldsymbol{H}_l^{\text{TGT}}\triangleq \boldsymbol{G}^H\text{diag}(\boldsymbol{h}^{\text{R-TGT}}_l)\in\mathbb{C}^{M \times N}\\
 & \boldsymbol{h}_{l,m}^{\text{TGT}}=(\boldsymbol{H}_l^{\text{TGT}}(m,:))^T \in\mathbb{C}^{N \times 1},
    \end{align}
\end{subequations}
where $k\in{\cal K}$, $m\in{\cal M}$, and $l\in{\cal L}$.

 To capture the complex non-linear correlations within the near-field CSI, the functions $f_0^{(0)}:\mathbb{R}^{2 (N+1) \times 1} \mapsto \mathbb{R}^{q / 2 \times 1}$ and $f^{(0)}_1:\mathbb{R}^{2N \times 1} \mapsto \mathbb{R}^{q / 2 \times 1}$ are implemented as two layer fully connected neural networks (FCNNs). Specifically, these functions process the CSI $\{\boldsymbol{h}_{k,m}^{\text{CU}}, \boldsymbol{h}_{l,m}^{\text{TGT}}\}$ to extract the features for pairs $(k, m)$ and $(l, m)$, respectively. 
 
 Denote by $\text{ReLU}(x)=(x)^+$ the rectified linear unit (ReLU) activation function. The feature vector from the BS antenna $m\in{\cal M}$ to CU $k\in{\cal K}$ is obtained as
\begin{align}
\nonumber
\boldsymbol{r}_{k,m}^{\text{CU}}&=f_0^{(0)}(\bar{\boldsymbol{h}}_{k,m}^\text{CU})\\
&=\boldsymbol{W}_{0,2}^{(0)} \left(\text{ReLU} \left( \boldsymbol{W}_{0,1}^{(0)} \bar{\boldsymbol{h}}_{k,m}^\text{CU} + \boldsymbol{b}_{0,1}^{(0)} \right)\right) + \boldsymbol{b}_{0,2}^{(0)},
\end{align}
where $\bar{\boldsymbol{h}}_{k,m}^\text{CU} =
[  (\Re \{ \boldsymbol{h}_{k,m}^{\text{CU}} \})^T , (\Im \{ \boldsymbol{h}_{k,m}^{\text{CU}} \} )^T]^{T}\in\mathbb{R}^{2(N+1) \times 1}$; $\boldsymbol{W}_{0,1}^{(0)} \in \mathbb{R}^{q \times 2(N+1)}$ and $\boldsymbol{W}_{0,2}^{(0)} \in \mathbb{R}^{q / 2 \times q}$ are the weight matrices for $f_0^{(0)}$; $\boldsymbol{b}_{0,1}^{(0)}\in \mathbb{R}^{q \times 1}$ and $\boldsymbol{b}_{0,2}^{(0)}\in \mathbb{R}^{q / 2 \times 1}$ denote the bias vectors for $f_0^{(0)}$. The feature vector from the BS antenna $m\in{\cal M}$ to TGT $l\in{L}$ is given by
\begin{align}
\boldsymbol{r}_{l,m}^{\text{TGT}}&=f^{(0)}_1(
\bar{\boldsymbol{h}}_{l,m}^{\text{TGT}})\notag \\
&=\boldsymbol{W}_{1,2}^{(0)} \left(\text{ReLU} \left( \boldsymbol{W}_{1,1}^{(0)} \bar{\boldsymbol{h}}_{l,m}^{\text{TGT}} + \boldsymbol{b}_{1,1}^{(0)} \right)\right) + \boldsymbol{b}_{1,2}^{(0)},
\end{align}
where $\bar{\boldsymbol{h}}_{l,m}^{\text{TGT}}=
[ (\Re \{ \boldsymbol{h}_{l,m}^{\text{TGT}}  \} )^T, (\Im \{ \boldsymbol{h}_{l,m}^{\text{TGT}}  \} )^T  ]^{T}\in\mathbb{R}^{2N \times 1}$; $\boldsymbol{W}_{1,1}^{(0)} \in \mathbb{R}^{q \times 2N}$ and $\boldsymbol{W}_{1,2}^{(0)} \in \mathbb{R}^{q / 2\times q}$ represent the weight matrices for $f_1^{(0)}$; $\boldsymbol{b}_{1,1}^{(0)}\in \mathbb{R}^{q \times 1},\boldsymbol{b}_{1,2}^{(0)}\in \mathbb{R}^{q / 2 \times 1}$ denote the bias vectors for $f_1^{(0)}$.

 First, with the obtained $\{\boldsymbol{r}_{k,m}^{\text{CU}},\boldsymbol{r}_{l,m}^{\text{TGT}}\}$, the feature information vectors $\boldsymbol{r}_{\text{CU}}^{(0)}$ and $\boldsymbol{r}_{\text{TGT}}^{(0)}$ are respectively given by 
\begin{subequations}
\begin{align}
\boldsymbol{r}_{\text{CU}}^{(0)}&=f_{\text{mean}}(\left \{\boldsymbol{r}_{k,m}^{\text{CU}}  \right \})= \frac{1}{KM}\sum_{k=1}^{K}\sum_{m=1}^M\boldsymbol{r}_{k,m}^{\text{CU}}\\
\boldsymbol{r}_{\text{TGT}}^{(0)}&=f_{\text{mean}}(\left \{\boldsymbol{r}_{l,m}^{\text{TGT}}  \right \})
=\frac{1}{LM}\sum_{l=1}^{L}\sum_{m=1}^M\boldsymbol{r}_{l,m}^{\text{TGT}},
\end{align}
\end{subequations}
where $f_{\text{mean}}$ denotes the arithmetic mean function in order to obtain CSI equally and retain the permutation invariance for the GNN model. 

Then, we obtain the initial feature vector $\boldsymbol{r}^{(0)}\in\mathbb{R}^{q \times 1}$ of RIS node via concatenating $\boldsymbol{r}_{\text{CU}}^{(0)}$ and $\boldsymbol{r}_{\text{TGT}}^{(0)}$, i.e.,
\begin{align}
\boldsymbol{r}^{(0)}=[(\boldsymbol{r}_{\text{CU}}^{(0)})^{T},(\boldsymbol{r}_{\text{TGT}}^{(0)})^{T}]^{T}\in \mathbb{R}^{q \times 1}.
\end{align}

Next, the initial feature vector $\boldsymbol{u}_k^{(0)}\in\mathbb{R}^{q  \times 1}$ for CU $k\in{\cal K}$ is generated by the FCNNs $f^{(0)}_{\text{CU}}:\mathbb{R}^{2 M(N+1) \times 1} \mapsto \mathbb{R}^{q \times 1}$, i.e.,
\begin{align}
\boldsymbol{u}_k^{(0)}&=f^{(0)}_{\text{CU}}(\bar{\boldsymbol{H}}_k^{\text{CU}})\\
&=\boldsymbol{W}_{\text{CU},2}^{(0)} \left(\text{ReLU} \left( \boldsymbol{W}_{\text{CU},1}^{(0)} \bar{\boldsymbol{H}}_k^{\text{CU}} + \boldsymbol{b}_{\text{CU},1}^{(0)} \right)\right) + \boldsymbol{b}_{\text{CU},2}^{(0)},\nonumber
\end{align}
where $\bar{\boldsymbol{H}}_k^{\text{CU}}=[ \text{vec}(\Re \{ \boldsymbol{H}_k^{\text{CU}} \} )^{T},\text{vec}(\Im \{ \boldsymbol{H}_k^{\text{CU}} \})^{T} ]^{T}\in\mathbb{R}^{2M(N+1) \times 1}$, $\boldsymbol{W}_{\text{CU},1}^{(0)} \in \mathbb{R}^{2q \times 2M(N+1)}$ and $\boldsymbol{W}_{\text{CU},2}^{(0)} \in \mathbb{R}^{q \times 2q}$ are the weight matrices, $\boldsymbol{b}_{\text{CU},1}^{(0)}\in \mathbb{R}^{2q \times 1}$ and $\boldsymbol{b}_{\text{CU},2}^{(0)}\in \mathbb{R}^{q \times 1}$ denote the bias vectors for $f^{(0)}_{\text{CU}}$.
 
Similarly, the initial feature vector $\boldsymbol{t}_l^{(0)}\in\mathbb{R}^{q  \times 1}$ for TGT $l\in{\cal L}$ is generated by the FCNNs $f^{(0)}_{\text{TGT}}:\mathbb{R}^{2 MN \times 1} \mapsto \mathbb{R}^{q \times 1}$. Therefore, we have
\begin{align}
\boldsymbol{t}_l^{(0)}&=f^{(0)}_{\text{TGT}}(\bar{\boldsymbol{H}}_l^{\text{TGT}})\\
&=\boldsymbol{W}_{\text{TGT},2}^{(0)} \left(\text{ReLU} \left( \boldsymbol{W}_{\text{TGT},1}^{(0)} \bar{\boldsymbol{H}}_l^{\text{TGT}} + \boldsymbol{b}_{\text{TGT},1}^{(0)} \right)\right) + \boldsymbol{b}_{\text{TGT},2}^{(0)},\nonumber
\end{align}
where $
\bar{\boldsymbol{H}}_l^{\text{TGT}}=[ \text{vec}(\Re\{ \boldsymbol{H}_l^{\text{TGT}} \} )^T,\text{vec}(\Im\{ \boldsymbol{H}_l^{\text{TGT}} \} )^{T}]^{T}\in\mathbb{R}^{2MN \times 1}$; $\boldsymbol{W}_{\text{TGT},1}^{(0)} \in \mathbb{R}^{2q \times 2MN}$ and $\boldsymbol{W}_{\text{TGT},2}^{(0)} \in \mathbb{R}^{q \times 2q}$ are the weight matrices, $\boldsymbol{b}_{\text{TGT},1}^{(0)}\in \mathbb{R}^{2q \times 1}$ and $\boldsymbol{b}_{\text{TGT},2}^{(0)}\in \mathbb{R}^{q \times 1}$ denote the bias vectors for $f^{(0)}_{\text{TGT}}$.

 \begin{figure*}[t]
    \centering
    \includegraphics[width=4.3in]{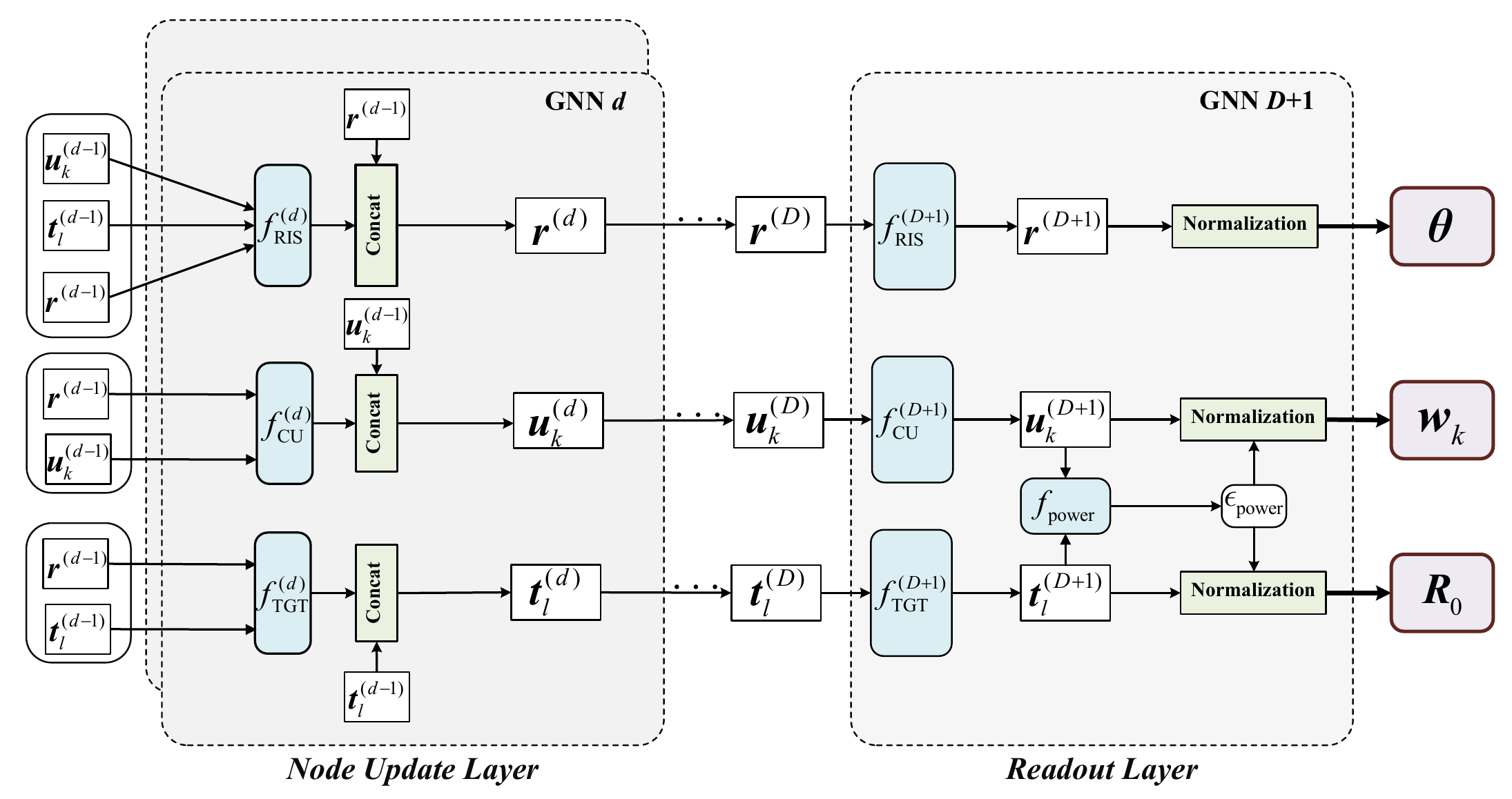}
    \caption{The structure of $d$-th node update layer, followed by a readout layer.}\label{fignode}
\end{figure*}

\subsubsection{Node Update Layers}
 In the node update layers of GNN model shown in Fig.~\ref{fignode}, it is required to iteratively update the feature vector for each node by aggregating and combining the information of its neighbors per node-update layer. Denote by $\boldsymbol{r}^{(d)}\in\mathbb{R}^{(d+1)q  \times 1}$, $\boldsymbol{u}_k^{(d)} \in\mathbb{R}^{(d+1)q  \times 1}$, and $\boldsymbol{t}_l^{(d)}\in\mathbb{R}^{(d+1)q \times 1}$  the feature vectors of RIS, CU $k\in{\cal K}$ and TGT $l\in{\cal L} $ updated at the $d$-th node update layer, where $d=1,...,D$. We have
\begin{equation}
\boldsymbol{r}^{(d)}=[(f^{(d)}_{\text{RIS}}(\boldsymbol{r}^{(d-1)},\boldsymbol{\varphi}_{\text{CU}}^{(d)},\boldsymbol{\varphi}_{\text{TGT}}^{(d)}))^T,(\boldsymbol{r}^{(d-1)})^T]^T,
\end{equation}
where $\boldsymbol{\varphi}_{\text{CU}}^{(d)}=\frac{1}{K}\sum_{k=1}^K \boldsymbol{u}_k^{(d-1)}$, $\boldsymbol{\varphi}_{\text{TGT}}^{(d)}=\frac{1}{L}\sum_{l=1}^L\boldsymbol{t}_l^{(d-1)}$ denote the element-wise arithmetic mean operations, and $f^{(d)}_{\text{RIS}}:\mathbb{R}^{dq\times dq \times dq} \mapsto \mathbb{R}^{q \times 1}$ is the FCNNs generating function for $\boldsymbol{r}^{(d)}$ at the $d$-th node-update layer. Accordingly, we have
\begin{align}
&f^{(d)}_{\text{RIS}}(\boldsymbol{r}^{(d-1)},\boldsymbol{\varphi}_{\text{CU}}^{(d)},\boldsymbol{\varphi}_{\text{TGT}}^{(d)}) =\boldsymbol{W}_{\text{RIS},2}^{(d)} \Big(\text{ReLU} \Big( \boldsymbol{W}_{\text{RIS},1}^{(d)} \bar{\boldsymbol{r}}^{(d-1)} \notag \\
&~~~~~~~~~~~~~ + \boldsymbol{b}_{\text{RIS},1}^{(d)} \Big)\Big) + \boldsymbol{b}_{\text{RIS},2}^{(d)},
\end{align}
where $\bar{\boldsymbol{r}}^{(d-1)}=[ (\boldsymbol{r}^{(d-1)})^T,(\boldsymbol{\varphi}_{\text{CU}}^{(d)})^T,(\boldsymbol{\varphi}_{\text{TGT}}^{(d)})^T]^{T}\in\mathbb{R}^{3dq \times 1}$; $\boldsymbol{W}_{\text{RIS},1}^{(d)} \in \mathbb{R}^{q \times 3dq}$ and $\boldsymbol{W}_{\text{RIS},2}^{(d)} \in \mathbb{R}^{q \times q}$ are the weight matrices; $\boldsymbol{b}_{\text{RIS},1}^{(d)}\in \mathbb{R}^{q \times 1}$ and $\boldsymbol{b}_{\text{RIS},2}^{(d)}\in \mathbb{R}^{q \times 1}$ denote the bias vectors for $f^{(d)}_{\text{RIS}}$.

Denote by $f^{(d)}_{\text{CU}}:\mathbb{R}^{dq \times dq \times dq} \mapsto \mathbb{R}^{q \times 1}$ and $f^{(d)}_{\text{TGT}}:\mathbb{R}^{dq\times dq\times dq} \mapsto \mathbb{R}^{q \times 1}$ the FCNNs for generating $\boldsymbol{u}_k^{(d)}$ and $\boldsymbol{t}_l^{(d)}$ of CU $k \in \cal K$ and TGT $l \in \cal L$ at the $d$-th layer, respectively, where $d=1,...,D$. It follows that
\begin{subequations}
\begin{align}
\boldsymbol{u}_k^{(d)}=[(f^{(d)}_{\text{CU}}(\boldsymbol{u}_k^{(d-1)},\boldsymbol{\varphi}_{\text{CU},k}^{(d)},\boldsymbol{r}^{(d-1)}))^T, (\boldsymbol{u}_k^{(d-1)})^T]^T
\\
\boldsymbol{t}_l^{(d)}=[(f^{(d)}_{\text{TGT}}(\boldsymbol{t}_l^{(d-1)},\boldsymbol{\varphi}_{\text{TGT},l}^{(d)},\boldsymbol{r}^{(d-1)}))^T, (\boldsymbol{t}_l^{(d-1)})^T]^T,
 \end{align}
\end{subequations}
where $k\in{\cal K}$, $l\in{L}$, and $d=1,...,D$. The functions $f^{(d)}_{\text{CU}}$ and $f^{(d)}_{\text{TGT}}$ are given by 
\begin{subequations}
\begin{align}
&f^{(d)}_{\text{CU}}(\boldsymbol{u}_k^{(d-1)},\boldsymbol{\varphi}_{\text{CU},k}^{(d)},\boldsymbol{r}^{(d-1)}) \notag \\
&=\boldsymbol{W}_{\text{CU},2}^{(d)} \left(\text{ReLU} \left( \boldsymbol{W}_{\text{CU},1}^{(d)} \bar{\boldsymbol{u}}_k^{(d-1)} + \boldsymbol{b}_{\text{CU},1}^{(d)} \right)\right) + \boldsymbol{b}_{\text{CU},2}^{(d)}\\
&f^{(d)}_{\text{TGT}}(\boldsymbol{t}_l^{(d-1)},\boldsymbol{\varphi}_{\text{TGT},l}^{(d)},\boldsymbol{r}^{(d-1)})\notag \\
&=\boldsymbol{W}_{\text{TGT},2}^{(d)} \left(\text{ReLU} \left( \boldsymbol{W}_{\text{TGT},1}^{(d)} \bar{\boldsymbol{t}}_l^{(d-1)} + \boldsymbol{b}_{\text{CU},1}^{(d)} \right)\right) + \boldsymbol{b}_{\text{CU},2}^{(d)},
\end{align}
\end{subequations}
where $\bar{\boldsymbol{u}}_k^{(d-1)}$ and $\bar{\boldsymbol{t}}_l^{(d-1)}$ are
\begin{subequations}
\begin{align}
\bar{\boldsymbol{u}}_k^{(d-1)}&=[ (\boldsymbol{u}_k^{(d-1)})^T,(\boldsymbol{\varphi}_{\text{CU},k}^{(d)})^T,(\boldsymbol{r}^{(d-1)})^T]^{T}\\
\bar{\boldsymbol{t}}_l^{(d-1)}&= [ (\boldsymbol{t}_l^{(d-1)})^T,(\boldsymbol{\varphi}_{\text{TGT},l}^{(d)})^T,(\boldsymbol{r}^{(d-1)})^T]^{T},
\end{align}
\end{subequations}
and $\boldsymbol{W}_{\text{CU},1}^{(d)} \in \mathbb{R}^{q \times 3dq}$ and $\boldsymbol{W}_{\text{CU},2}^{(d)} \in \mathbb{R}^{q \times q}$ are the weight matrices for $f^{(d)}_{\text{CU}}$; $\boldsymbol{W}_{\text{TGT},1}^{(d)} \in \mathbb{R}^{q \times 3dq}$ and $\boldsymbol{W}_{\text{TGT},2}^{(d)} \in \mathbb{R}^{q \times q}$ represent the weight matrices for $f^{(d)}_{\text{TGT}}$; $\boldsymbol{b}_{\text{CU},1}^{(d)}\in \mathbb{R}^{q \times 1}$ and $\boldsymbol{b}_{\text{CU},2}^{(d)}\in \mathbb{R}^{q \times 1}$ denote the bias vectors for $f^{(d)}_{\text{CU}}$; $\boldsymbol{b}_{\text{TGT},1}^{(d)}\in \mathbb{R}^{q \times 1}$ and $\boldsymbol{b}_{\text{TGT},2}^{(d)}\in \mathbb{R}^{q \times 1}$ denote the bias vectors for $f^{(d)}_{\text{TGT}}$. 

In addition, we have $\boldsymbol{\varphi}_{\text{CU},k}^{(d)}=f_{\max}(\{\boldsymbol{u}_j^{(d-1)}\}_{j\neq k})$ and $\boldsymbol{\varphi}_{\text{TGT},l}^{(d)}=f_{\max}(\{\boldsymbol{t}_j^{(d-1)}\}_{j\neq l})$ with $f_{\max}$ denoting the element-wise max-pooling function, which performs well empirically and corresponds to the fact that interference is mainly dominated by the strongest user. Therefore, we have 
\begin{subequations}
\begin{align}
[\boldsymbol{\varphi}_{\text{CU},k}^{(d)}]_i=\max\{[\boldsymbol{u}_j^{(d-1)}]_i\mid j\in{\cal K}\setminus\{k\}\}\\
[\boldsymbol{\varphi}_{\text{TGT},l}^{(d)}]_i=\max\{[\boldsymbol{t}_j^{(d-1)}]_i\mid j\in{\cal L}\setminus\{l\}\},
\end{align}
\end{subequations}
where $i=1,...,dq$.

After a number of $D$ node update layers, we obtain the final feature vectors $\{\boldsymbol{r}^{(D)}, \boldsymbol{u}_k^{(D)}, \boldsymbol{t}_l^{(D)}\}$ of the GNN nodes.

\subsubsection{Readout Layer}
 Fig.~\ref{fignode} shows the readout layer of the proposed GNN model. As the output of the GNN readout layer with linearizing and normalizing the final feature vectors of the nodes, the real and imaginary components of the BS's ISAC transmit beamforming design variables $(\{\boldsymbol{w}_k\}_{k=1}^K, \boldsymbol{R}_0)$ and XL-RIS's reflecting coefficient vector $\boldsymbol{\theta}$ is finally generated. For notational convenience, the $(D+1)$-th layer is used to denote the readout layer.

 For the RIS node, we first employ the FCNNs $f^{(D+1)}_{\text{RIS}}:\mathbb{R}^{(D+1)q \times 1} \mapsto \mathbb{R}^{2N \times 1}$ to linearize $\boldsymbol{r}^{(D)}\in\mathbb{R}^{2N  \times 1}$. Accordingly, we have
\begin{align}
\boldsymbol{r}^{(D+1)}&=f^{(D+1)}_{\text{RIS}}(\boldsymbol{r}^{(D)})= \boldsymbol{W}_{\text{RIS},1}^{(D+1)} \boldsymbol{r}^{(D)} + \boldsymbol{b}_{\text{RIS},1}^{(D+1)},
\end{align}
where $\boldsymbol{W}_{\text{RIS},1}^{(D+1)}\in\mathbb{R}^{ 2N\times (D+1)q}$ and $\boldsymbol{b}_{\text{RIS},1}^{(D+1)}\in\mathbb{R}^{ 2N\times 1}$ denote the weight matrix and bias vector for $f^{(D+1)}_{\text{RIS}}$.

 Then the normalization operation for guaranteeing the modulus-one constraint \eqref{eq.p12} is used to obtain the XL-RIS reflecting coefficient vector $\boldsymbol{\theta}$, i.e.,
\begin{equation}\label{thetaGNN}
[\boldsymbol{\theta}]_{n}=\frac{[\boldsymbol{r}^{(D+1)}]_{n}}{\bar{r}_{n}} + j\frac{[\boldsymbol{r}^{(D+1)}]_{N+n}}{\bar{r}_{n}},~n=1,...,N,
\end{equation}
where $\bar{r}_{n}=\sqrt{\left([\boldsymbol{r}^{(D+1)}]_{n}\right)^{2}+\left([\boldsymbol{r}^{(D+1)}]_{N+n}\right)^{2}}$. Once with $\boldsymbol{\theta}$, we readily obtain the XL-RIS reflecting coefficient matrix $\boldsymbol{\Phi}$.

To adaptively balance the BS transmit power allocation between communication and sensing  while strictly satisfying the constraint \eqref{eq.p11}, we first introduce the learnable power-split factor $\epsilon_{\text{power}}\in (0, 1)$, which is denoted by the function $f_{\text{power}}:\mathbb{R}^{(D+1)q \times (D+1)q} \mapsto \mathbb{R}^{2M \times 1}$. We then have 
\begin{subequations}\label{power}
\begin{align}
\epsilon_{\text{power}}&=f_{\text{power}}(\boldsymbol{u}_k^{(D+1)},\boldsymbol{t}^{(D+1)})
\\&=\text{Sigmoid}\left(\boldsymbol{W}_{\text{power}}^T\boldsymbol{\varphi}_{\text{power}}+b_{\text{power}} \right),
\end{align}
\end{subequations}
where $\boldsymbol{\varphi}_{\text{power}}= \frac{1}{K+1} \left ( \sum_{k \in \cal K} \boldsymbol{u}_k^{(D+1)}+\boldsymbol{t}^{(D+1)} \right ) \in\mathbb{R}^{ 2M\times1}$; $\text{Sigmoid}(x)=\frac{1}{1+e^{-x}} $ denotes the Sigmoid activation function; $\boldsymbol{W}_{\text{power}}\in\mathbb{R}^{ 2M\times1}$, and $b_{\text{power}}\in\mathbb{R}$ are the two power-split parameters in $f_{\text{power}}$. The feature vectors of CU $k\in{\cal K}$ and TGT $l\in{\cal L}$ at readout layer, $\boldsymbol{u}_k^{(D+1)}\in\mathbb{R}^{2M\times 1}$ and $\boldsymbol{t}^{(D+1)}\in\mathbb{R}^{2M\times 1}$, are given by
\begin{subequations}\label{ukD+1}
\begin{align}
\boldsymbol{u}_k^{(D+1)}&=f^{(D+1)}_{\text{CU}}(\boldsymbol{u}_k^{(D)})=\boldsymbol{W}_{\text{CU},1}^{(D+1)} \boldsymbol{u}_k^{(D)} + \boldsymbol{b}_{\text{CU},1}^{(D+1)}\\
\boldsymbol{t}^{(D+1)}&=f^{(D+1)}_{\text{TGT}}(\frac{1}{L}\sum_{l=1}^L\boldsymbol{t}_l^{(D)})\notag \\
&=\boldsymbol{W}_{\text{TGT},1}^{(D+1)} (\frac{1}{L}\sum_{l=1}^L\boldsymbol{t}_l^{(D)}) + \boldsymbol{b}_{\text{TGT},1}^{(D+1)},
\end{align}
\end{subequations}
where $\boldsymbol{W}_{\text{CU},1}^{(D+1)}\in\mathbb{R}^{ 2M\times (D+1)q}$ and $\boldsymbol{b}_{\text{CU},1}^{(D+1)}\in\mathbb{R}^{ 2M\times 1}$ denote the weight matrix and bias vector for the linearizing function $f^{(D+1)}_{\text{CU}}:\mathbb{R}^{(D+1)q \times 1} \mapsto \mathbb{R}^{2M \times 1}$. Similarly for sensing $L$ targets, the feature vector $\boldsymbol{t}^{(D+1)}\in\mathbb{R}^{2M  \times 1}$ containing the information of $L$ targets is generated by the linear function $f^{(D+1)}_{\text{TGT}}:\mathbb{R}^{(D+1)q \times 1} \mapsto \mathbb{R}^{2M \times 1}$ with $\boldsymbol{W}_{\text{TGT},1}^{(D+1)}\in\mathbb{R}^{ 2M\times (D+1)q}$ and $\boldsymbol{b}_{\text{TGT},1}^{(D+1)}\in\mathbb{R}^{ 2M\times (D+1)q}$.   

Accordingly, the BS transmit beamforming vector $\boldsymbol{w}_k$ for each CU $k \in \cal K$ is obtained as 
\begin{align}\label{wk_GNN}  [{\boldsymbol{w}}_{k}]_m= \alpha_0^{\text{CU}}([\boldsymbol{u}_k^{(D+1)}]_m+j[\boldsymbol{u}_k^{(D+1)}]_{M+m}),
\end{align}
where $m\in{\cal M}$, and $\alpha_0^{\text{CU}}=\frac{\sqrt{\epsilon_{\text{power}}P_0}}{\sum_{k=1}^{K}\left\|\boldsymbol{u}_k^{(D+1)}\right\|^2}$ denotes the power scaling factor, $m \in{\cal M}$. With ${\boldsymbol{w}}_{k}$ at hand, we are ready to construct the BS communication beamforming matrix $\boldsymbol{W}$.

For obtaining the BS sensing covariance matrix $\boldsymbol{R}_0$, we have 
\begin{equation}\label{R0_GNN}
[\boldsymbol{R}_0]_{m,m}=\alpha_0^{\text{TGT}}\left(([\boldsymbol{t}^{(D+1)}]_m)^2+([\boldsymbol{t}^{(D+1)}]_{M+m})^2\right),
\end{equation}
where $m\in{\cal M}$, and $\alpha_0^{\text{TGT}}=\frac{(1-\epsilon_{\text{power}})P_0}{\text{tr}(\boldsymbol{R}_0)}$ denotes the power scaling factor. Note that in \eqref{wk_GNN} and \eqref{R0_GNN} that the normalization ensures $\text{tr}(\boldsymbol{W}^H\boldsymbol{W})=\epsilon_{\text{power}}P_0$ and $\text{tr}(\boldsymbol{R}_0)= (1-\epsilon_{\text{power}})P_0$, which satisfies the BS transmit power constraint \eqref{eq.p11}.

% by using $\boldsymbol{t}_l^{(D)}$ as the input to , which is given by

% \begin{equation}\label{tlD+1}
% \boldsymbol{t}_l^{(D+1)}=f^{(D+1)}_{\text{target}}(\boldsymbol{t}_l^{(D)})
%  \in\mathbb{R}^{2M  \times 1},
% \end{equation}
% then the sensing covariance matrix $\boldsymbol{R}_0$ is obtained as follows

% \begin{equation}\label{R0}
% [\boldsymbol{R}_0]_{m,m}=[\boldsymbol{t}_l^{(D+1)}]_m+j[\boldsymbol{t}_l^{(D+1)}]_{M+m},\forall m \in{\cal M},
% \end{equation}

% Note that the constraints \eqref{eq.p11} and \eqref{eq.p12} are guaranteed to be satisfied by proper normalization and quantization in the readout layer.

\subsection{GNN Training Procedure and Analysis}
The negative value of the weighted sum rate of $K$ CUs plus the penalty of sensing beampatterns for $L$ TGTs is treated as the loss function $f_{loss}$. The loss function is given by
\begin{equation}\label{loss}
f_{loss}(\boldsymbol{W},\boldsymbol{R}_0,\boldsymbol{\theta}) = -\sum_{k=1}^{K} \tau_k\log _{2}\left(1+\gamma_{k}\right) + \beta \sum_{l=1}^{L} (\rho_{th} - \rho_l)^+,\notag
\end{equation}
where $\beta>0$ denotes the parameter of the penalty term $\sum_{l=1}^{L}(\rho_{th}-\rho_l)^+$. During the GNN training process, our objective is to minimize the loss function $f_{loss}$ so as to maximize the achievable weight sum rate of all CUs while satisfying the sensing constraints in \eqref{eq.p15d}. 

Then, the unsupervised training procedure with the loss function is shown in Algorithm~2, where $E$, $B$, and $B_{\text{size}}$ denote the numbers of epoch, batch, and batch sizes in the training dataset $\cal D$; $\omega$ and $\Theta_{\text{GNN}}$ denote the learning rate and the parameters in the proposed GNN, respectively.

 The computational complexity of the proposed GNN-based method is analyzed as follows, which is primarily determined by feature initialization, node update, and variables readout.

\begin{itemize}
    \item \emph{Feature Initialization}: The channel matrices $\boldsymbol{H}_k^{\text{CU}}\in\mathbb{C}^{M  \times (N+1)}$, $\forall k\in{\cal K}$, and $\boldsymbol{H}_l^{\text{TGT}}\in\mathbb{C}^{M  \times N}$, $\forall l\in{\cal L}$, are mapped to a feature space of dimension $q$ via FCNNs. The complexity is given by $\mathcal{O}((K+L)MNq)$.
    \item \emph{Node Update}: In each of $D$ layers, the aggregation and update operations involve matrix multiplications with the feature dimension $q$. The complexity for updating all nodes is $\mathcal{O}(D(K+L)q^2)$.
    \item \emph{Variable Readout}: The transformation in the readout layer maps the features back to the optimization variables, with a complexity of $\mathcal{O}(K(M+N)q)$.
\end{itemize}

\begin{algorithm}[h]
\caption{Proposed Unsupervised Training Procedure for (P1) based on GNN}
\label{alg:GNN_training}
\begin{algorithmic}[1]
\REQUIRE Training dataset $\mathcal{D}$ containing CSI samples $\{\boldsymbol{H}_k^{\text{CU}},\boldsymbol{H}_l^{\text{TGT}}\}$; Batch size $B$; Number of epoch $E$; Learning rate $\omega$.
\ENSURE Optimized GNN parameters $\Theta_{\text{GNN}}$ (weights and biases) to obtain optimal $\boldsymbol{\theta},  (\boldsymbol{W},\boldsymbol{R}_0)$
\STATE Initialize GNN parameters $\Theta_{\text{GNN}}$ randomly.
\FOR{epoch $e = 1, \dots, E$}
    \FOR{batch $b = 1, \dots, B$}
        \STATE \textbf{Step 1: Initializing and Updating in Initial Layer and $D$ Node Update Layers}
        \STATE Input a batch of $B_{\text{size}}$ CSI samples into the GNN to obtain the initial features $\boldsymbol{r}^{(0)}$, $\boldsymbol{u}_k^{(0)}$ and $\boldsymbol{t}_l^{(0)}$ 
        \STATE Obtaining node features $u_k^{(D)}$ and $t^{(D)}$ by passing $D$ node update layers
        \STATE \textbf{Step 2: Variable Recovery in Readout Layer}
        \STATE Recover XL-RIS reflecting coefficients $\boldsymbol{\theta}$ via \eqref{thetaGNN} with unit-modulus normalization.
        \STATE Compute power-split factor $\epsilon_{\text{power}}$ via \eqref{power}.
        \STATE Recover $\boldsymbol{w}_k$ via \eqref{wk_GNN} with power scaling.
        \STATE Recover sensing covariance matrix $\boldsymbol{R}_0$ via \eqref{R0_GNN} with power scaling.

        \STATE \textbf{Step 3: Loss Computation}
        \STATE Calculate the SINR $\gamma_k$ for each CU $k\in \cal K$ and sensing gain $\rho_l$ for each TGT $l\in \cal L$, and compute the loss function $f_{loss}$. 
         
         % according to \eqref{loss}.
%         \begin{equation*}
% f_{loss}= -\sum_{k=1}^{K} \tau_k\log _{2}\left(1+\gamma_{k}\right) + \beta \sum_{l=1}^{L} (\rho_{th} - \rho_l)^+.
%         \end{equation*}
        
        \STATE \textbf{Step 4: Backpropagation}
        \STATE Compute gradients $\nabla_{\Theta_{\text{GNN}}}f_{loss}$ for $f_{loss}$.
        \STATE Update GNN parameters $\Theta_{\text{GNN}}$ using Adam optimizer: $\Theta_{\text{GNN}} \leftarrow \Theta_{\text{GNN}} - \omega \nabla_{\Theta_{\text{GNN}}}f_{loss} $.
    \ENDFOR
\ENDFOR
\RETURN The trained GNN model parameters $\Theta_{\text{GNN}}$.
\end{algorithmic}
\end{algorithm}

%\vspace{-0.2cm}
\section{Numerical Results}\label{5}
In this section, numerical results are provided to evaluate the performance of the proposed GNN model. An evaluation method for the proposed GNN is introduced based on several metrics including four aspects: computational efficiency, feasibility, robustness, and the ability of generalization.

\subsection{Simulation Setup}

\subsubsection{Simulation Scenario}
The path loss exponent $\alpha$ are set to be 2.5, 2.2, 3.5, and 2.5 for the XL-RIS $\to$ CU, BS $\to$ XL-RIS, BS $\to$ CU, and XL-RIS $\to$ TGT channels, respectively. Unless otherwise stated, we set $(M, K, L)=(9, 4, 2)$, the BS's transmit power budget and path-loss coefficient are set to $P_0=1$ W and $\rho_0=\frac{\lambda}{4\pi}$, respectively. The number of XL-RIS elements is set to be $N \in \left \{ 21^2,23^2,25^2,27^2,29^2,31^2 \right \}$, where $N_x=N_z$. The carrier frequency of BS transmit signal is set to $f_c=28$ GHz (i.e., the wavelength is $\lambda=1.07$ cm). The sensing beampattern gain threshold and receiver noise power of CU $k$ are set as $\rho_{th}=10^4$ and $\sigma_k^2=1\times10^{-9}$, respectively. $\tau_k$ is set to $1$. We consider a coordinate setup measured in meters, where the center of BS and XL-RIS are $(-4\mathrm{~m},2\mathrm{~m},0\mathrm{~m})$ and $(0\mathrm{~m},0\mathrm{~m},0\mathrm{~m})$, respectively. Each CU $k\in \cal K$ is randomly dropped within a circle $\cal C_{\text{CU}}$ centered at $(2\mathrm{~m},2\mathrm{~m},0\mathrm{~m})$ with a radius of $R_{\text{CU}}=1\mathrm{~m}$, and each TGT $l\in\cal L$ is randomly distributed within a circle $\cal C_{\text{TGT}}$ centered at $(4\mathrm{~m},2\mathrm{~m},0\mathrm{~m})$ with a radius of $R_{\text{TGT}}=1\mathrm{~m}$.

% \begin{itemize}
%     \item \textbf{Coordinate setup for $K$ CUs:} Each CU $k\in \cal K$ is randomly dropped within a circle $\cal C_{\text{CU}}$ centered at $(2\mathrm{~m},2\mathrm{~m},0\mathrm{~m})$ with a radius of $R_{\text{CU}}=1\mathrm{~m}$.
%     \item \textbf{Coordinate setup for $L$ TGTs:} Each TGT $l\in\cal L$ is randomly distributed within a circle $\cal C_{\text{TGT}}$ centered at $(4\mathrm{~m},2\mathrm{~m},0\mathrm{~m})$ with a radius of $R_{\text{TGT}}=1\mathrm{~m}$.
% \end{itemize}

\subsubsection{Implementation Details}
The proposed GNN-based optimization framework is implemented using Python 3.8.18 and Pytorch 1.13.1 with CUDA 11.7, whereas the proposed BCD method is solved via the CVX toolbox for $(\boldsymbol{W},\boldsymbol{R}_0)$ and the Manopt solver for $\boldsymbol{\theta}$ in MATLAB R2024b. For the neural network training process, the Adaptive Moment Estimation (Adam) optimizer is employed with the learning rate of $\omega = 10^{-3}$. The numbers of epoch, batch, and batch size are set as $E=2000$, $B=4$, and $B_{\text{size}}=128$, respectively. The learnable power-splitting factor $\epsilon_{power}$ is initialized to $0.7$. All numerical simulations are conducted on a server with one NVIDIA RTX 3090 GPU (24 GB of memory).

\subsubsection{Training Dataset}
The dataset $\cal{D}$ for GNN training procedure, consisting of CSI samples without labels, is randomly generated based on the geometric relationships (i.e., the locations of $K$ CUs and $L$ TGTs) in order to ensure exhaustive coverage of diverse physical scenarios and channel realizations.

\subsection{Proposed GNN Structure}
The parameters of the GNN model are set as $D \in \{3,4\}$, $q=128$, and $\beta=200$. The first two parameters are designed to enhance the feature extraction, while the relatively large regularization parameter ensures the satisfaction of constraint \eqref{eq.p15d}. Furthermore, to assess the robustness of our proposed GNN model, the imperfect CSI of $\boldsymbol{h}^{\text{R-CU}}_k$ and $\boldsymbol{h}^{\text{BS-CU}}_k$ with imperfect parameter $\sigma_e^2 \in \{0.001,0.01\}$ is introduced during training process ($\sigma_e^2=0$ denotes the perfect CSI), while using perfect CSI for computing sum rate.

For performance comparison, we consider the following two baseline schemes for the near-field ISAC system designs.

\begin{itemize}
    \item \textbf{DNN-based design scheme:} The layer-number and feature-dimension parameters of the proposed DNN model are set as $D\in \{3,4\}$ and $q=128$, respectively, and the regularization parameter in the DNN loss function is set as $\beta=200$. The DNN consists of one initial layer, $D$ node update layers and one readout layer, similar to GNN, which generates $\{\boldsymbol{w}_k\}_{k=1}^K$, $\boldsymbol{R}_0$, and $\boldsymbol{\theta}$ through the readout layer.
    \item \textbf{BCD-based design scheme:} For the BCD-based algorithm for (P1), parameters $(C_1,C_2,C)$ are set to be $(50,50,30)$. Finally, the tolerance level is set to be $\epsilon=10^{-6}$ in the proposed Algorithm~1. The maximum number of iterations $I_{max}$ is set to $30$. 
\end{itemize}

\begin{figure}[h]
    \centering
    \includegraphics[width=3in]{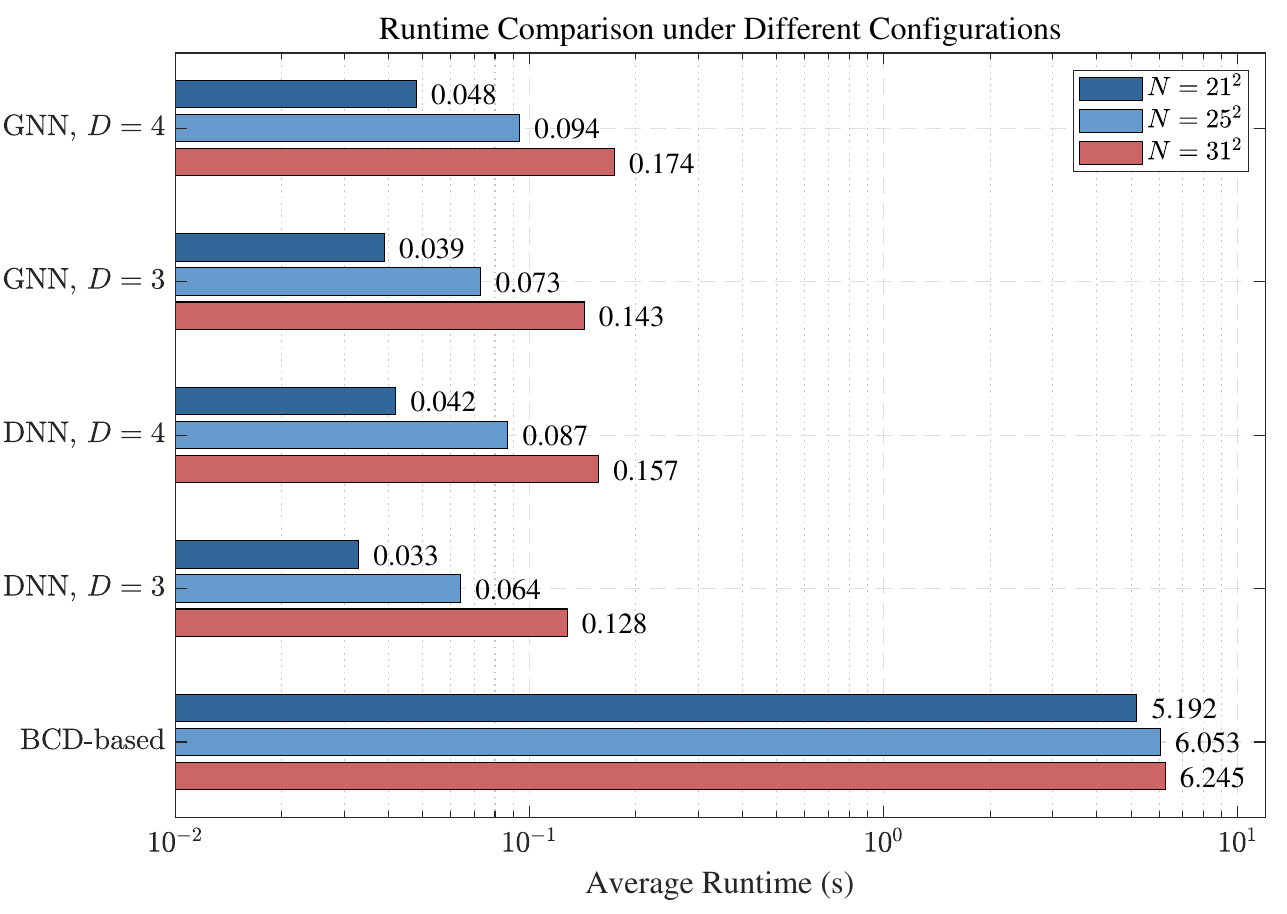}
    \caption{Runtime Comparison under Different Configurations (Unit: Seconds)}\label{figtime}
\end{figure}

We first illustrate the impact of the number of node update layer $D$ on the WSR in Fig.~\ref{figtime}, which compares the average runtime of the proposed GNN model against the DNN and BCD-based benchmark schemes across different $N$. As illustrated, the traditional BCD-based algorithm incurs a significantly high computational latency. In contrast, learning-based methods (GNN and DNN) demonstrate remarkable speed, highlighting its potential for real-time ISAC applications. Compared to DNN, the slightly increased latency of GNN is attributed to the message-passing mechanism in node update layers, which is, however, essential for capturing the spatial topology of ISAC systems effectively.

\renewcommand{\arraystretch}{1.25}
\begin{table}[h]
  \centering
  \caption{The Satisfaction Rate of Constraint ($16\textnormal{d}$) with XL-RIS Reflecting Element Number $N$}
  \label{tab:cons rate}
  \begin{tabular}{|c|c|c|c|c|c|c|c|}
    \hline
    $(K, L)=(4, 2)$   & $N = 21^2$      & $N = 25^2$      & $N = 31^2$   \\ 
    \hline
    GNN, $D=3,\sigma_e^2=0$               & 99.9\%    & 99.9\%    & 99.9\%  \\
    \hline
    GNN, $D=3,\sigma_e^2=0.001$                & 99.5\%    & 99.6\%    & 99.7\%  \\
    \hline
    GNN, $D=3,\sigma_e^2=0.01$                 & 99.2\%  & 99.3\%  & 99.5\% \\
    \hline
    GNN, $D=4,\sigma_e^2=0.001$                 & 98.9\%  & 99.1\%  & 99.1\% \\
    \hline
    GNN, $D=4,\sigma_e^2=0.01$                 & 98.4\%  & 98.7\%  & 98.9\% \\
    \hline
  \end{tabular}
\end{table}

Table~\ref{tab:cons rate} illustrates the sensing performance (i.e., satisfaction rate of constraint \eqref{eq.p15d} with different values of $\sigma_e^2$ and $D$), which shows the robustness of the GNN for ISAC system, i.e., the satisfaction rate of constraint changes only slightly with different values of $\sigma_e^2$. Furthermore, the sensing performance of $3$-layer GNN is better than $4$-layer GNN.

\begin{figure}[h]
    \centering
    \includegraphics[width=2.6in]{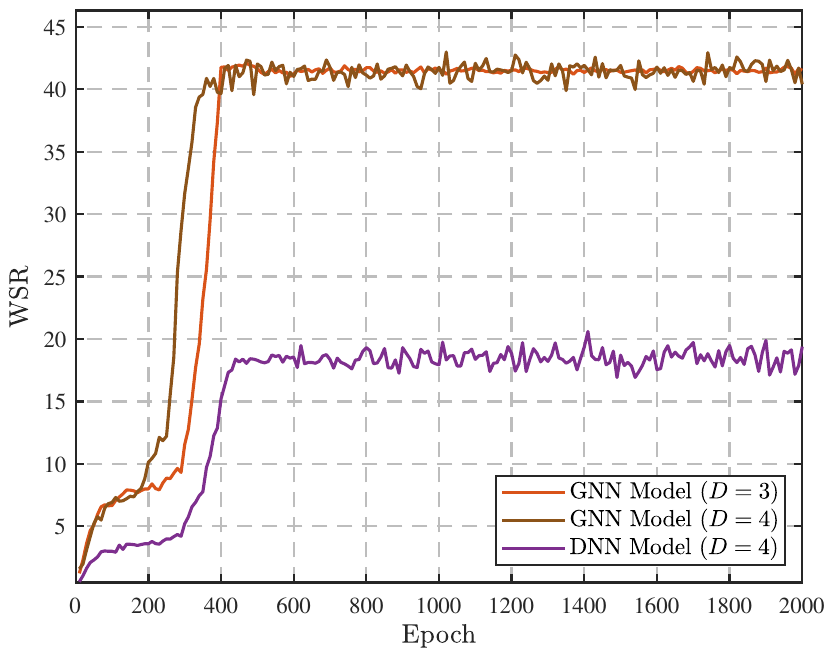}
    \caption{Convergence behavior of the proposed GNN model and DNN for the ISAC system with $N=21^2$.}\label{figcon}
\end{figure}

\begin{figure}[h]
    \centering
    \includegraphics[width=2.6in]{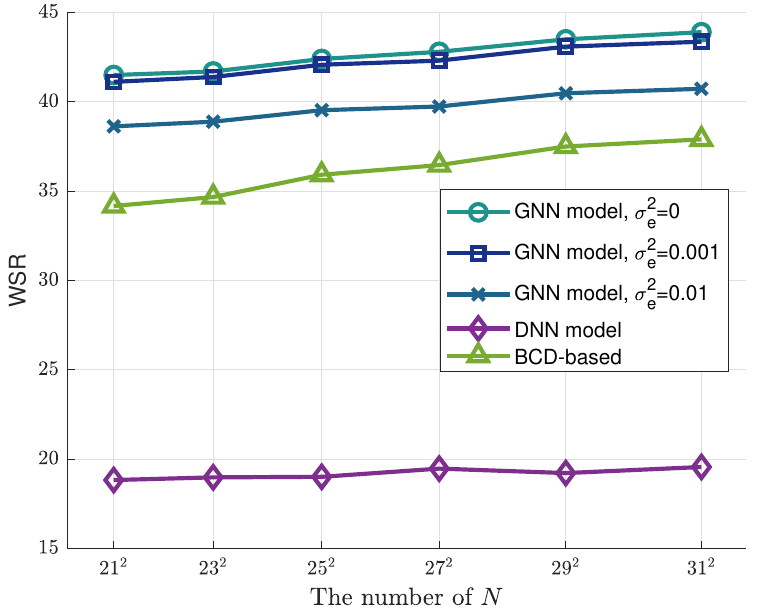}
    \caption{The performance of WSR versus the number of $N$.}\label{fig3}
\end{figure}

\begin{figure}[h]
    \centering
    \includegraphics[width=2.6in]{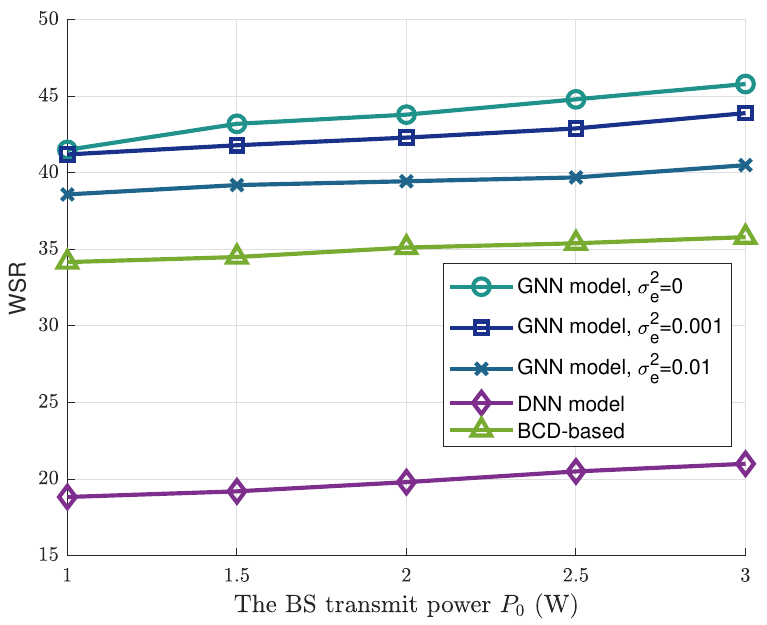}
    \caption{The performance of WSR versus the BS transmit power $P_0$ with $N=21^2$.}\label{figpower}
\end{figure}

\begin{figure}[h]
    \centering
    \includegraphics[width=2.6in]{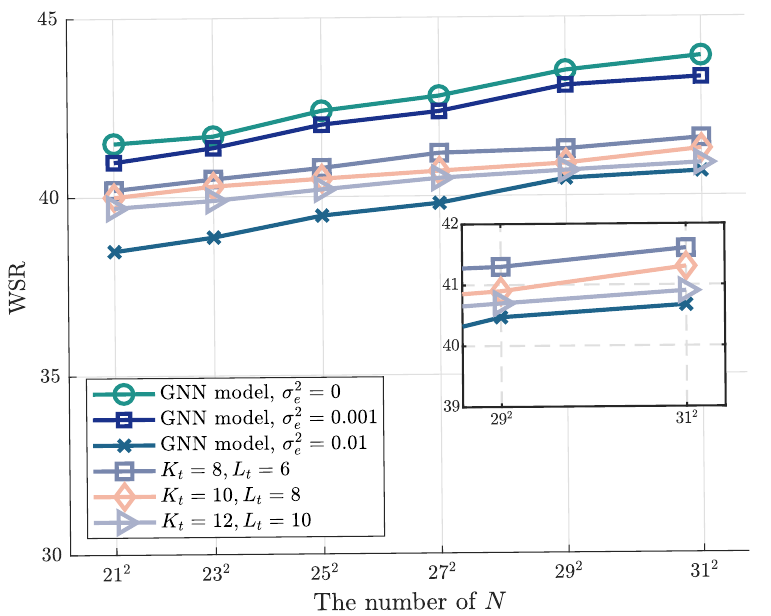}
    \caption{The generalizability of the proposed GNN model trained with $(K, L)=(4, 2)$.}\label{figm1}
\end{figure}

 Fig.~\ref{figcon} shows the convergence behavior of the proposed GNN model with different number of node update layers, $D$. It can be observed that the GNN model with $D=3$ achieves a similar and more stable WSR compared to the case with $D=4$ upon convergence, indicating that $D=3$ is sufficient for the aggregation and combination in the near-field ISAC than $D=4$. Accordingly, $3$-layer GNN achieves the balance between feature extraction capability and training stability compared to $4$-layer GNN, which suffers from a higher variance during training due to the over-smoothing inherent in GNN. Therefore, $D=3$ is adopted for the subsequent simulations in the near-field ISAC.

\begin{figure}[h]
    \centering
    \includegraphics[width=2.6in]{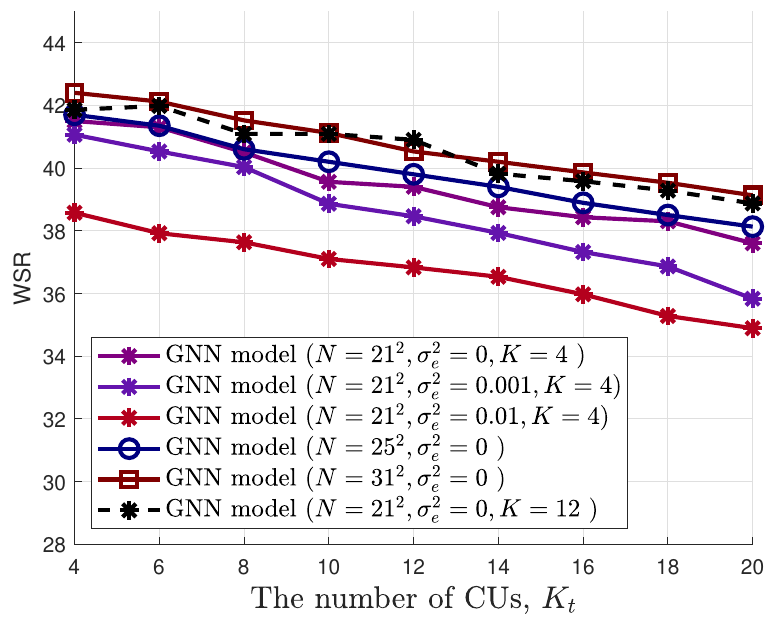}
    \caption{The performance of WSR versus the number of $K_t$ in the proposed GNN model trained with $K$.}\label{figm2}
\end{figure}

Fig.~\ref{fig3} shows the communication performance comparison of different schemes. The proposed GNN achieves good performance over the BCD-based algorithm, the inferior performance of DNN compared to GNN and BCD is attributed to its inherent limitation in capturing the complex topological dependencies among the XL-RIS, CUs, and TGT. Unlike GNN, conventional DNN model treats the CSIs as unstructured vectors, thereby failing to fully exploit the spatial correlations in the high-complexity near-field channels, struggling with feature extraction of CSI, and limiting their performance in near-field ISAC. Furthermore, the proposed GNN shows the robustness of communication performance with imperfect parameter $\sigma_e^2 \in \{0.001,0.01\}$, i.e., the WSR changes only slightly with different values of $\sigma_e^2$. Furthermore, Fig.~\ref{figpower} illustrates the performance of WSR with $P_0$. 
 
 Due to the permutation invariance of GNN, we next study the generalizability of the proposed GNN model trained with smaller $(K,L)=(4,2)$ by evaluating its performance with the numbers of CUs and TGTs larger than $K$ and $L$, i.e., $K_{t}$ and $L_{t}$. As expected in Fig.~\ref{figm1}, the proposed GNN based design solution demonstrates the remarkable flexibility in serving different ISAC systems, while maintaining a good communication performance (whereas DNN requires retraining). From another perspective, we consider the generalizability of the GNN model trained with $K\in \{4,12\}$. Fig.~\ref{figm2} demonstrates the communication performance of the GNN model for different values of $K_{t}$. The WSR decreases with an increase of $K_{t}$, while still preserving good communication performance. For GNN model trained with $K=12$, a critical insight can be observed regarding the degree of WSR degradation. Specifically, the model trained with $K=12$ exhibits a significantly slower decline in WSR as $K_t>12$, compared to the model trained with $K=4$. This indicates that selecting a training set with a moderate number of CUs enables the GNN to learn more complex interference management patterns, thereby enhancing its robustness and mitigating the performance loss when generalizing to another near-field ISAC.

%\vspace{-0.2cm}
\section{Conclusion}\label{6}
In this paper, we studied a joint beamforming design problem where the objective is to maximize the WSR for XL-RIS assisted near-field ISAC systems, while ensuring the BS transmit power and sensing beampattern gain constraints, as well as the modulus constraints on the XL-RIS phase shift. Using the FP reformulation and SCA method, we developed a BCD-based algorithm to obtain a near-optimal solution. Further, to address the dynamic topology caused by CU/TGT mobility and the complex coupled near-field channel models for ISAC, while overcoming the scalability issues of conventional DL approaches, we proposed a GNN-based joint beamforming design framework for near-field ISAC. The near-field ISAC system was modeled as a heterogeneous graph comprising XL-RIS/CU/TGT nodes to effectively exploit near-field CSI. The message passing mechanism is employed to enable effective information exchange among directly connected nodes, and each XL-RIS/CU/TGT node is associated with a feature vector, which can be effectively mapped to the BS transmit beamforming variables or XL-RIS reflecting coefficient vector. Numerical results showed the superiority of the proposed GNN based scheme, in terms of efficiency, feasibility, robustness, and generalizability, compared to the existing benchmark schemes.

\vspace{12pt}

\end{document}